\documentclass[%
superscriptaddress,
twocolumn,
 amsmath,amssymb,
 aps,
prb,
]{revtex4-2}

\usepackage{graphicx}
\usepackage{dcolumn}
\usepackage{bm}
\usepackage{amsmath}
\usepackage{epsfig}
\usepackage{amssymb}
\usepackage{bbm}
\usepackage{braket}
\usepackage{float}
\usepackage{enumitem}
\usepackage{tensor}
\usepackage{xcolor}
\usepackage{siunitx}
\usepackage[colorlinks=true,citecolor=blue,linkcolor=blue,urlcolor=blue]{hyperref}

\usepackage{color}

\begin{document}


\title{Cavity modification of magnetoplasmon mode through coupling with intersubband polaritons}

\author{Lucy L. Hale}
\affiliation{
 ETH Z\"urich, Institute of Quantum Electronics,
	Auguste-Piccard-Hof 1,
    8093 Z\"urich, Switzerland
}%
\email{luhale@phys.ethz.ch}
\author{Daniele De Bernardis}
\affiliation{National Institute of Optics (CNR-INO), c/o LENS via Nello Carrara 1, Sesto F.no 50019, Italy}
\author{Stephan Lempereur}%
\affiliation{
 ETH Z\"urich, Institute of Quantum Electronics,
	Auguste-Piccard-Hof 1,
    8093 Z\"urich, Switzerland
}%
\author{Lianhe H. Li}
\affiliation{School of Electronic and Electrical Engineering, University of Leeds, Leeds LS2 9JT, United Kingdom}
\author{A. Giles Davies} 
\affiliation{School of Electronic and Electrical Engineering, University of Leeds, Leeds LS2 9JT, United Kingdom}
\author{Edmund H. Linfield}
\affiliation{School of Electronic and Electrical Engineering, University of Leeds, Leeds LS2 9JT, United Kingdom}
\author{Trevor Blaikie}
\affiliation{Department of Electrical and Computer Engineering, University of Waterloo, 200 University Ave W, Waterloo, Ontario N2L 3G1, Canada}
\author{Chris Deimert}
\affiliation{Department of Electrical and Computer Engineering, University of Waterloo, 200 University Ave W, Waterloo, Ontario N2L 3G1, Canada}
\author{Zbigniew R. Wasilewski}
\affiliation{Department of Electrical and Computer Engineering, University of Waterloo, 200 University Ave W, Waterloo, Ontario N2L 3G1, Canada}
\author{Iacopo Carusotto}
 \affiliation{
    Pitaevskii BEC Center, CNR-INO and Dipartimento di Fisica, Universit\`a di Trento, I-38123 Trento, Italy
}%
\author{Jean-Michel Manceau}
\affiliation{Centre de Nanosciences et de Nanotechnologies (C2N),
	CNRS UMR 9001, University of Paris-Saclay,
	Palaiseau 91120, France
}%
\author{Mathieu Jeannin}
\affiliation{Centre de Nanosciences et de Nanotechnologies (C2N),
	CNRS UMR 9001, University of Paris-Saclay,
	Palaiseau 91120, France
}%
\author{Raffaele Colombelli}
\affiliation{Centre de Nanosciences et de Nanotechnologies (C2N),
	CNRS UMR 9001, University of Paris-Saclay,
	Palaiseau 91120, France
}%
\author{J\'er\^ome Faist}%
\affiliation{
 ETH Z\"urich, Institute of Quantum Electronics,
	Auguste-Piccard-Hof 1,
    8093 Z\"urich, Switzerland
}%
\author{Giacomo Scalari}
\affiliation{
 ETH Z\"urich, Institute of Quantum Electronics,
	Auguste-Piccard-Hof 1,
    8093 Z\"urich, Switzerland
}%


\date{\today}

\begin{abstract} 

We investigate the coupling of a multi-mode metal-insulator-metal cavity to a two-dimensional electron gas (2DEG) in a quantum well in the presence of a strong magnetic field.
The TM cavity mode is strongly hybridized with an intersubband transition of the 2DEG, forming a polaritonic mode in the ultrastrong coupling regime, while the TE mode remains an almost purely cavity mode.
The magnetoplasmon excitation emerging from the presence of the magnetic field couples with both TM and TE modes, exhibiting different coupling strengths and 
levels of spatial field inhomogeneity. While the strong homogeneity of the bare TE mode gives rise to the standard anticrossing of strong coupling, the inhomogeneous polaritonic TM mode is shown to activate an observable Coulombic effect in the spectral response, often referred to as non-locality.
This experiment demonstrates a cavity-induced modification of the 2DEG response and offers a new route to probing the effect of Coulomb interactions in ultrastrongly coupled systems via reshaping of their cavity mode profiles. 

\end{abstract}

\maketitle

\section{\label{sec:level1}Introduction} 

Quantum wells (QWs) embedded in terahertz (THz) metal-insulator-metal (MIM) cavities are a powerful platform for exploring strong light–matter interactions, with high coupling strengths enabled by large dipole moments, collective enhancement, and deeply subwavelength field confinement \cite{FriskKockum2019, solano_RevModPhys.91.025005, Scalari2012, Ciuti2005, DeBernardis:24}. 
Typically, the cavity modes are coupled to QW intersubband transitions (ISBTs), where the coupling strength can reach the ultrastrong coupling regime \cite{Todorov2010,Geiser_PhysRevLett.108.106402,Dietze_2013,Askenazi_2014,Jeannin_2019,Goulain2023,Todorov2009,Berkmann2024} characterized by a light–matter interaction strength comparable to the bare resonance frequencies. 
This so-called ultrastrong coupling regime holds the promise of bringing quantum effects and technologies into the far-infrared range of the spectrum. Examples include the predicted emission of correlated photon pairs through abrupt modulation of the system \cite{Ciuti2005}, or the electrical probing of quantum phase transitions \cite{iqbal_PhysRevResearch.6.033097}.

Along a different line, significant efforts have focused on exploiting the bosonic nature of ISB polaritons to develop inversionless lasers \cite{deliberato_PhysRevLett.102.136403}. Notable demonstrations include ISB polariton–LO phonon scattering \cite{Delteil_PhysRevB.83.081404, Manceau_10.1063/1.5029893}, and more recently, final-state stimulation in polariton–polariton scattering schemes \cite{Knorr2022}.
With an eye towards applications, there has also been a surge of activity aimed at toggling the system in and out of the strong coupling regime via electrical or optical modulation. Such dynamic control is particularly valuable for implementing amplitude or phase modulation on a continuous-wave (CW) carrier at GHz frequencies, as demonstrated in \cite{Anappara_10.1063/1.2006976, Pirotta2021, Chung_https://doi.org/10.1002/advs.202207520, Malerba_10.1063/5.0213965}, and has been recently proposed as a means to realize ultrafast saturable absorber mirrors \cite{Jeannin2021, Jeannin2023}.

While the ISB-MIM cavity approach relies on fixed cavity and material configurations to engineer the light–matter coupling, magnetic field–induced cyclotron resonances offer dynamic, in-situ tunability \cite{Scalari2012, Rajabali2021}. Landau polaritons are formed by strongly coupling light to the collective cyclotron resonance – known as a magnetoplasmon (MP). These systems have enabled the highest recorded coupling strengths, as well as allowing direct tuning of the frequency the magnetoplasmon with applied magnetic field \cite{Bayer2017,Mornhinweg2024}. 
\begin{figure*}
	\centering
    \includegraphics[width = \textwidth]{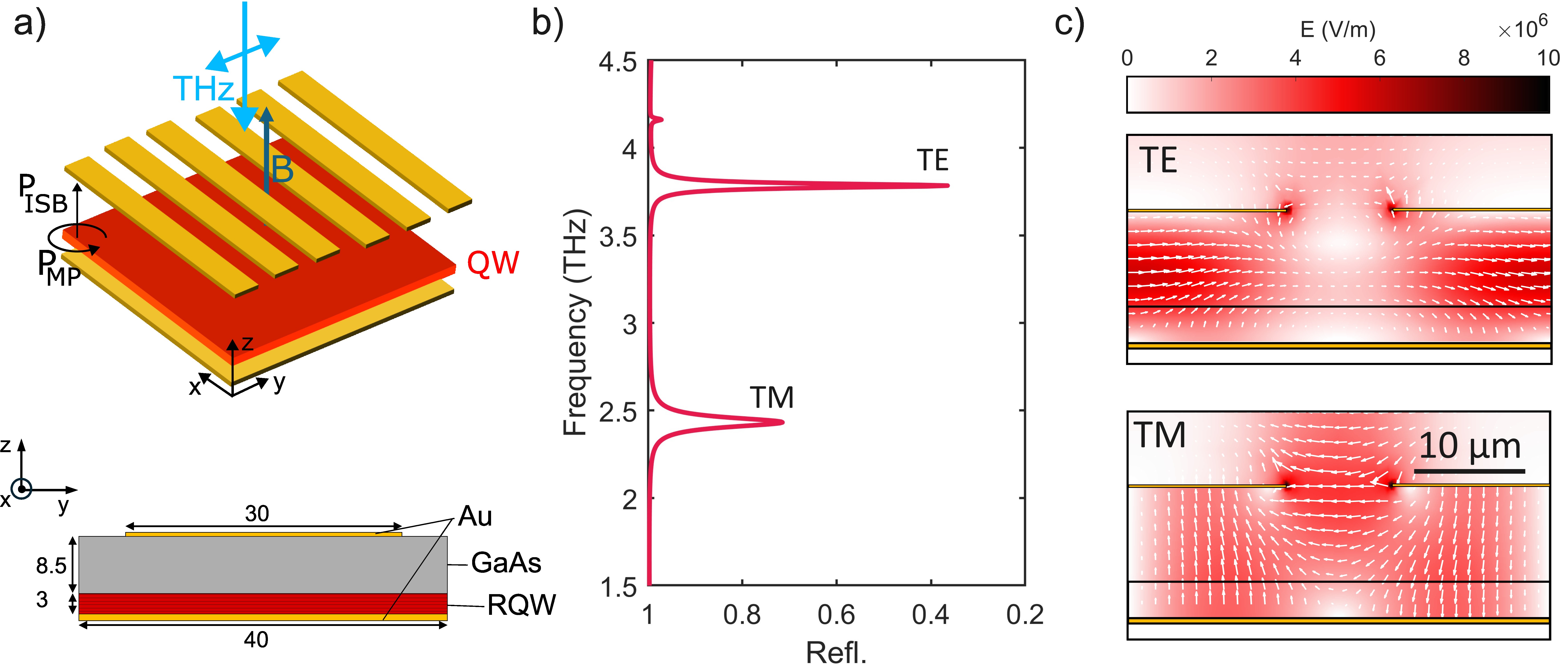}
	\caption{Metal-Insulator-Metal Cavities: a) Schematic showing the sample, indicating multi quantum well layer (red), magnetic (B) field direction, incident THz field and intersubband ($P_{ISB}$) and MP  ($P_{MP}$) resonances in the QW. Below: side profile of cavity showing dimensions in $\mathrm{{\mu}m}$. b). Finite-element simulation of the reflectance spectra of cavity in the absence of QW, indicating the TM and TE modes. In the following of the work, both modes couple to the MP resonance; in addition, the TM mode is strongly coupled to the ISB resonance.  c) Simulated in-plane electric field in a cavity cross-section at frequencies corresponding to the TM and TE modes. Color map shows electric field magnitude in the y-z plane, white arrows show the in-plane field orientation.}
	\label{fig:fig1}
\end{figure*}

In Landau polariton systems, the description of the coupling relies on Kohn’s theorem, which states that the cyclotron resonance frequency in a translationally-invariant two-dimensional gas is unaffected by electron-electron interactions \cite{Kohn_PhysRev.123.1242}. 
Despite its simplicity, this result turns out to be extremely powerful, reducing the description of the system to a non-interacting one, and allowing for the development of a simple, intuitive picture of light-matter interactions in terms of a Hopfield-like polaritonic description \cite{hopefield_PhysRev.112.1555,ciuti_PhysRevB.81.235303}. 
At the same time, Kohn's theorem is also a limitation, excluding the possibility to observe non-linearities and possible many-body physics in the optical properties of these systems.
Going beyond its range of validity is thus becoming an important topic of research \cite{Monticone_2025}, which turns out to be surprisingly challenging:
besides the use of intense THz sources as in Ref. \cite{Maag_break_kohn_nature2016}, the fundamental strategy to break Kohn's theorem and observe the effect of Coulomb interactions in the cyclotron resonance is to design the system to explicitly break translational invariance.
This can be naturally achieved in resonant cavities with highly sub-wavelength field confinement and inhomogeneous fields \cite{keller_few_electrons_2017}, where the violation of Kohn's theorem is manifested as a non-local polaritonic response \cite{Rajabali2021, Rajabali:23, endo2025cavitymediatedcouplinglocalnonlocal}.

In this work, we demonstrate the breakdown of Kohn's theorem in a strongly hybridized cavity-electron system, comprising of quantum wells strongly coupled to a metal–insulator–metal (MIM) cavity. 
In contrast to conventional MIM polariton systems, where the cavity couples only to the intersubband transition (ISBT), we apply a magnetic field, activating also the magnetoplasmon (MP) resonance of the two-dimensional electron gas, in a tripartite MIM-ISBT-MP interplay.

Using THz time-domain spectroscopy, we observe a pronounced renormalization of the magnetoplasmon resonance frequency through electron–electron Coulomb interactions. This is activated due to the spatial inhomogeneity of the TM cavity mode, which is strongly hybridized with the ISB transition. Unlike Ref.~\cite{Rajabali2021}, where non-locality resulted from a deeply sub-wavelength cavity miniaturization design ($\lambda/1000$), here the violation of Kohn’s theorem occurs with moderate field confinement ($\lambda/60$), provided that the resonant mode is sufficiently inhomogeneous.
Furthermore, we demonstrate that by tuning the magnetoplasmon with the magnetic field, it is possible to strongly couple it with a different TE mode, which is mostly homogeneous and in-plane polarized.
As a consequence, the TE mode is not hybridized with the ISBT, and thus exhibits purely cavity-like behavior.
Moreover, the non-locality disappears and Kohn's theorem is re-established due to a high level of spatial homogeneity.

This device thus allows sampling of the MP degree of non-locality by sweeping the magnetic field and by resonantly selecting cavity/polaritonic modes with varying homogeneity, largely increasing the flexibility of previous standard sub-wavelength cavity miniaturization designs~\cite{keller_few_electrons_2017,Rajabali2021}.
Interestingly, the resonant control on the non-local response of strongly hybridized cavity-electron systems is purely cavity-induced, effectively realizing a semi-classical counterpart of the modification of materials induced by the cavity quantum vacuum \cite{ciuti_review,Enkner2025}. 

\begin{figure*}
	\centering
    \includegraphics[width = \textwidth]{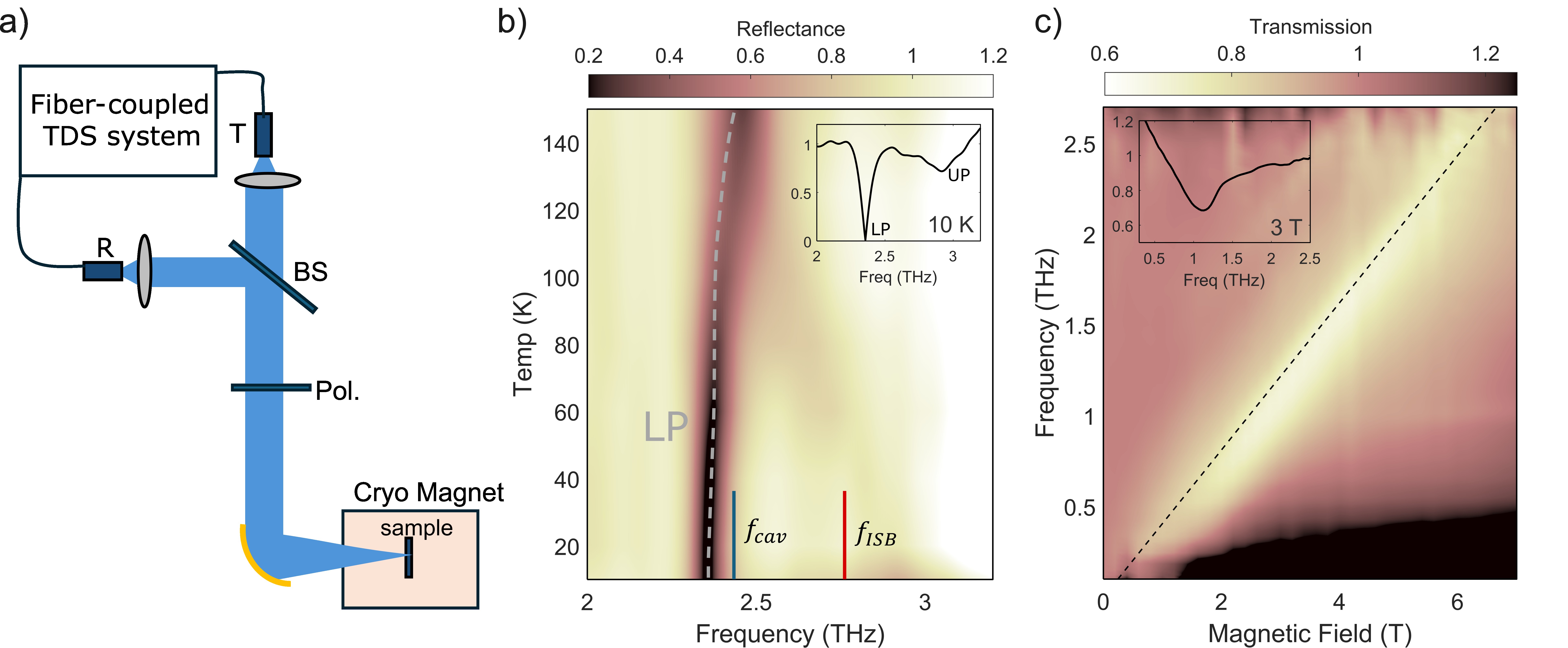}
	\caption{a) Schematic of experimental reflection THz-TDS set up. THz transmitter and receiver indicated with a `T' and `R' respectively. b) Measured reflectance spectra of the QW-loaded cavity as a function of temperature (with no applied magnetic field). The lower polariton branch (LP) is indicated. Below schematic indicating the frequency mismatch between the cavity and ISB resonances. Inset - single spectra slice at 10 K. c) Measured THz-TDS transmission spectra of semiconductor heterostructure in the absence of a cavity at 3K. The black dashed line indicates the unperturbed MP frequency given by $\omega_B=eB/m^*$, where $m^* = 0.067m_e$. Inset - single spectra slice at 3T. Regions in (b) and (c) where reflectance and transmission are greater than 1.0 are unphysical and due to normalization to reference traces with low signal. However, these are included for clarity of the data.}
	\label{fig:fig2}
\end{figure*}
\section{\label{sec:level2}Experimental Set-Up}
\label{Methods}  

\subsection{Cavity composition}\label{cavity}
The MIM cavity design is shown in Figure \ref{fig:fig1}a. The QWs (red) are sandwiched between a gold back plane and a gold grating with a periodicity of 40 $\mathrm{{\mu}m}$ and a duty cycle of 75\%. The cavity supports a variety of modes when excited at normal incidence with THz light polarized perpendicular to the grating (shown in Fig. \ref{fig:fig1}a). The first order transverse-magnetic (TM) and transverse-electric (TE) modes are shown in Fig. \ref{fig:fig1}b,c. The TM mode sits at 2.4 THz and exhibits a strong electric field amplitude perpendicular to the growth plane (z-direction in \ref{fig:fig1}a) underneath the gold grating. In the gaps, however, the TM mode also supports considerable in-plane fields (y-direction in \ref{fig:fig1}a).  In addition, the TE mode at 3.8 THz also lies within the spectral bandwidth of our measurement system. This has primarily in-plane fields, which are strongest underneath the gold grating. 

\subsection{Quantum Well}
A \SI{3}{\micro\meter} stack of $N_{\rm qw}=$ 53 GaAs/AlGaAs rectangular quantum wells (RQWs) with an electron density per well of $3 \times 10^{11} \mathrm{cm^{-2}}$ is placed in the cavity below an \SI{8}{\micro\meter} GaAs spacer. The QWs exhibit two resonances in the THz region which we investigate in our study: the ISBT and the MP resonance.

The ISBT is at 2.74 THz and strongly couples to the TM cavity mode due to its strong z-polarized fields. Figure \ref{fig:fig2}b shows the THz-TDS reflection spectra measured as a function of temperature in the absence of a magnetic field (experimental set-up schematic shown in Fig. \ref{fig:fig2}a). At 150 K, a single spectral feature around 2.4 THz is resolved, corresponding to the TM cavity resonance. As the temperature is reduced, the ISBT becomes well-resolved and couples to the TM mode, resulting in the formation of polariton branches. The difference in spectral contrast observed between upper and lower polariton branches is a result of the mismatch in frequency between the cavity TM resonance (2.4 THz) and the ISBT (2.74 THz).

When a magnetic field is applied, the QW also exhibits an MP resonance due to the collective cyclotron resonance of the electrons \cite{KWChiu1974}. Unlike the ISBT, which is fixed in frequency with the material growth, the MP resonance frequency increases linearly with applied magnetic field as $\omega_B=eB/m^*$, where $m^* = 0.067m_e$ is the effective mass of electrons in the GaAs QW.
The MP resonance in the bare QW heterostructure without the cavity is measured using transmission THz-TDS is shown in Fig. \ref{fig:fig2}c.
The THz response shows an almost perfect agreement with the expected linear dependence on the magnetic field (shown as a dashed black line). The linewidth can be estimated to have an almost $B$-independent  value of $\kappa/(2\pi)\approx 300\,$GHz (see inset), which is the expected order of magnitude considering the number of quantum wells and electron density \cite{Zhang2014}. We will see later that this clear agreement of the observed MP frequency with the expected $\omega_B$ is in contrast to the cavity case, where the cavity modes play a prominent role in reshaping the MP optical response.

\section{\label{sec:level3}THz Magnetospectroscopy of Strongly Coupled Cavity }

\begin{figure*}
	\centering
    \includegraphics[width = 15cm]{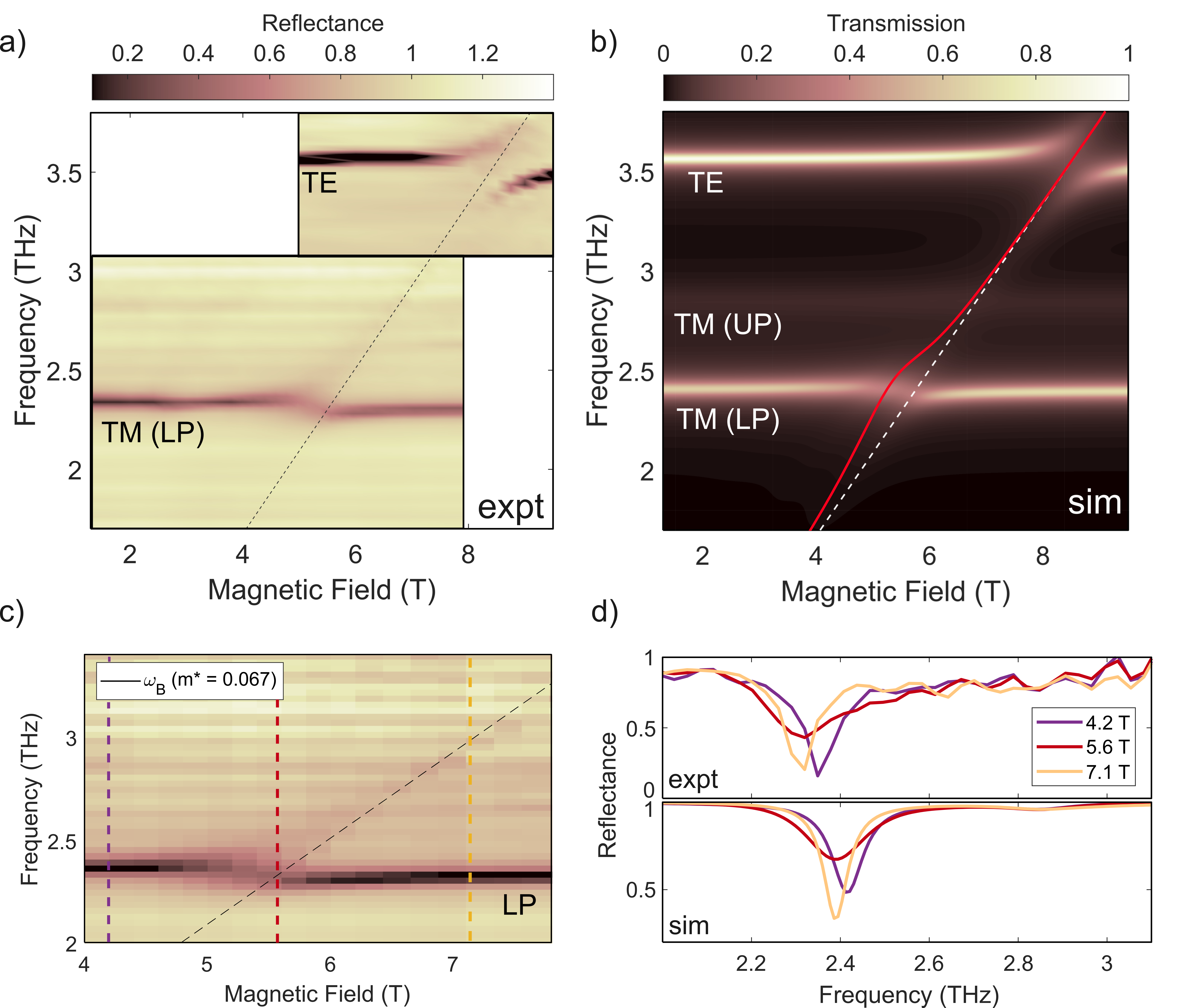}
	\caption{Interaction of cavity modes with MP resonance through magnetic field tuning: a). Measured reflectance spectra at 3K as a function of magnetic field strength. Dashed line indicates the cyclotron frequency, $\omega_B$. TM (LP) indicates the lower polariton branch of the ISB polariton with the TM mode, TE indicates the TE mode. b) Calculated spectra using the formalism in Section \ref{sec:coulomb}, accounting for Coulombic interaction. For the theory parameters, see Table \ref{tab:numerical_params} in App. \ref{sec:params}. Both TM upper (UP) and lower (LP) polaritons are visible, as well as TE mode. The white dashed line is the bare cyclotron $\omega_B$, while the red solid line is a guide for the eye, obtained by interpolating between the bare cyclotron frequency and the brightest, blue-shifted, cyclotron k-mode (see Appendix \ref{app:3modes_trans} for further details). c) Close-up of non-smoothed spectra of lower polariton branch. In (a,c), the black dashed line indicates the bare MP resonance as illustrated in Fig. \ref{fig:fig2}(c).
d) Individual spectra of TM LP for three values of the magnetic field corresponding to $\omega_B < \omega_{TM}$ (purple), $\omega_B = \omega_{TM}$ (red) and $\omega_B > \omega_{TM}$ (yellow).}
	\label{fig:figspectra}
\end{figure*}

The cavity-coupled QW device is then measured by reflection-mode THz-TDS at 3K with an applied DC magnetic field from 0 – 9 T. In this magnetic field range, the MP resonance sweeps across the entire spectral range of the measurement and interacts with the different cavity modes for different values of the B field. The resulting spectra are shown in Figure \ref{fig:figspectra}a.  The MP resonance frequency, given by $\omega_B =eB/m^{*}$, is shown by the dashed line. For both the TE and TM modes, we observe a clear coupling of the cavity mode to the MP; however, the nature of the coupling is different depending on the mode.

The high contrast feature observed in the spectra at 2.4 THz corresponds to the lower ISB polariton, where the TM cavity mode is coupled to the ISBT. The much lower contrast upper polariton is less visible in the spectra, and we therefore restrict our analysis to the lower polariton only.  Although the TM mode is primarily polarized orthogonally to the MP, it still supports sufficient in-plane field components to couple to the MP. The observed coupling around 4 - 6 T (shown more clearly in Fig. \ref{fig:figspectra}c,d) therefore corresponds to a tripartite coupling between the ISBT and cavity (which form the ISB polariton) with the MP. 
This LP-MP coupling does not result in the typical anti-crossing shape, but rather in an enhanced broadening of the spectral cavity feature due to its mixing with the broadened MP resonance. 
Most interestingly, the location of this broadening feature is displaced from the crossing point with the bare MP frequency $\omega_B$, as seen most clearly in the interpolated color map of Fig. \ref{fig:figspectra}a. This can be understood as resulting from a significant blue-shift of the MP which is the signature of the non-local effects stemming from the Coulomb interaction, as highlighted by the red line in the simulation panel in Fig. \ref{fig:figspectra}b. The origin of this blue-shift will be discussed in more detail in Section \ref{sec:discussion}. 

On the other hand, the higher-frequency TE mode at 3.5 THz is a purely photonic mode that has no contribution from the ISBT. 
It contains mostly in-plane fields which directly couple to the MP resonance, resulting in strong coupling and a clear anti-crossing behaviour. 
In contrast to the spectral region around the TM mode, here the MP appears to fit well the $\omega_B$ dependency on magnetic field. Fitting the response using a Hopfield model (assuming an MP frequency $\omega_B$) results in a polariton gap of 0.6 THz and therefore normalized coupling of $\eta = 0.04$. 

The physical understanding of the different coupling of the TM and TE mode to the MP and ISBT resonances in the QW will be discussed in the following Section \ref{sec:discussion}.

\section{\label{sec:discussion}{Discussion}}

\subsection{\label{sec:coulomb}{Cavity-induced Coulomb interactions}}

In order to fully understand the interactions between the cavity modes and the MP - particularly where this deviates from what is expected - it is necessary to consider the impact of the spatially-resolved inhomogeneous fields inside the cavity.

From the theory of macroscopic dielectrics, it is well known that a polarizable medium is strongly affected by its own internal electric field.
This self-interaction is due to the fact that the dipolar constituents of the material interact among each other via the Coulomb force, which is often approximated by the dipole-dipole interaction.
When the medium is excited by a radiative source, it develops a macroscopic polarization density vector $\textbf{P}(\textbf{r}, t)$, oscillating together with the external drive. 
This macroscopic motion of charges also gives rise to a macroscopic electric field inside the dielectric, which splits into bulk and surface contributions \cite{griffiths2013introduction}:
\begin{equation}\label{eq:E-field_inside}
\begin{split}
    & \textbf{E}_{\rm inside}(\textbf{r}, t) = \textbf{E}_{\rm bulk}(\textbf{r}, t) + \textbf{E}_{\rm surf}(\textbf{r}, t)
    \\
    = & \frac{(\textbf{r}-\textbf{r}')}{4\pi\epsilon_0} \Bigg[ \int_V d^3r'\, \frac{\vec{\nabla}'\cdot\textbf{P}(\textbf{r}',t)}{|\textbf{r}-\textbf{r}'|^3} -\int_{\partial V} d\textbf{S}'\cdot \frac{\textbf{P}(\textbf{r}', t)}{|\textbf{r}-\textbf{r}'|^3} \Bigg],
\end{split}
\end{equation}

The energy contribution due to this internal electric field, $-\textbf{P}\cdot \textbf{E}_{\rm inside}$ , which is dependent on the dielectric geometry, makes it harder to polarize the medium and thus leads to what is called \emph{depolarization shift} \cite{Todorov2010,todorovPhysRevB.85.045304,todorovPhysRevB.91.125409,de_bernardis_cavity_2018}.
In most situations, the system is so large that the surface contribution becomes negligible, and the only remaining contribution comes from the inhomogeneous bulk term.

Using a classical coarse-grained approach \cite{jackson_classical_2013} (as described in detail in the Appendix) the inclusion of such a depolarization effect results in a $\textbf{k}$-dependent (wavevector dependent) MP frequency in the 2D QW:

\begin{equation}\label{eq:omB_dispersion}
    \bar{\omega}_B^2(\textbf{k})\approx \omega_B^2 + N_{\rm qw}\omega_P^2\zeta_{\textbf{k}}.
\end{equation}
 with an explicit dependence on the material's plasma frequency
\begin{equation}\label{eq:plasma_freq}
    \omega_P = \sqrt{\frac{e^2 n_{2D}}{\epsilon m L_c}}\,.
\end{equation}
Here $N_{\rm qw}$ is the number of quantum wells, and the function $\zeta_{\textbf{k}}$ depends on the specific cavity-QW geometry. 
Following App. \ref{sec:Green}, for the simplest cavity geometry, with two infinite metallic parallel plates at distance $L_c$, we obtain the long-wavelength limit: 
\begin{equation}\label{eq:longwave_exp_zetak}
    \zeta_{\textbf{k}}\approx \frac{L_ck}{2}-\frac{(L_c k)^2}{4},
\end{equation}
holding for $L_ck\ll 1$.
The first term of Eq. \eqref{eq:longwave_exp_zetak} accounts for the free space contribution of Coulomb interactions, while the second term, proportional to $(L_ck)^2$, is the screening correction coming from the metallic boundaries (not to be confused with the surface contribution in Eq. \eqref{eq:E-field_inside}, which refers only to the QW dielectric properties, while the metallic boundary affects the $(\textbf{r}-\textbf{r}')/|\textbf{r}-\textbf{r}'|^3$ kernel in Eq. \eqref{eq:E-field_inside}, being replaced with the proper Green's function [see Appendix \ref{sec:Green}]).
This phenomenology is also commonly known as \emph{non-locality} \cite{Rajabali2021}.

Without inhomogeneities in a translationally invariant geometry, the MP is excited only on the $\textbf{k}=0$ mode so 
the system does not experience any depolarization shift. In this case the system is in agreement with Kohn's theorem and is well described by a completely non-interacting system, without any trace of its internal Coulomb forces. However, in our observed configuration, the cavity-ISBT polariton excites the quantum well with a highly inhomogeneous electric field profile (due to its TM mode structure), leading to a non-homogeneous MP oscillating polarization $\vec{\nabla}\cdot \textbf{P}\neq 0$.
The cavity thus induces a visible Coulombic contribution in the MP dynamics.
\begin{figure*}
	\centering
    \includegraphics[width = \textwidth]{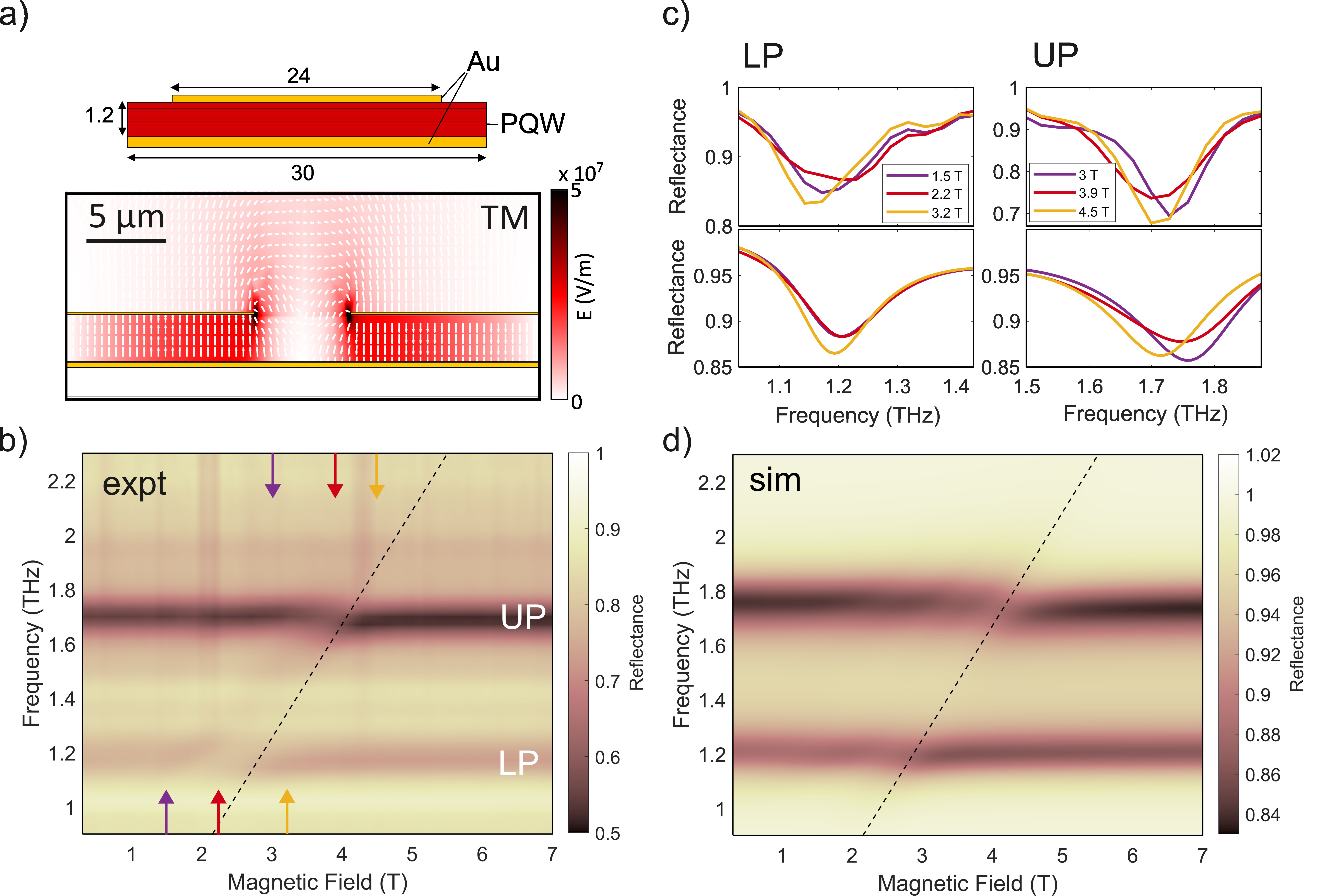}
	\caption{a) Cavity containing parabolic quantum well heterostructure: schematic (above) and simulated cavity fields (below) b) Experimentally measured spectra with magnetic field for cavity in a. Arrows indicate positions of line cuts in (c) for lower polariton (bottom) and upper polariton (top). c) Individual spectra of TM LP (left) and UP (right) for three values of the magnetic field corresponding to the purple, red and yellow arrows in Fig. 4a.  d) Calculated transmission of sample in b. For the theory parameters, see Table \ref{tab:numerical_params} in App. \ref{sec:params}.}
	\label{fig:parabolic}
\end{figure*}
\subsection{\label{sec:modelling}Modeling the Coupled System}

Tuning the magnetic field directly selects the level of inhomogeneity in the MIM cavity through resonant excitation of the different modes coupled to the MP. 
The choice of cavity mode in light-matter coupled systems is therefore important not only for maximizing the strength of the coupling with the electronic transition, but also for determining the degree of inhomogeneity and, consequently, the breaking of translational invariance of the system. 

To demonstrate this, in Figure \ref{fig:figspectra}b we calculate the spectra taking into account the effect of Coulomb interaction and resulting wavelength-dependent MP frequency. 
This calculation is performed with a simple classical linear model, starting from Maxwell equations in the presence of a polarizable material and projecting them on the cavity two-mode subspace (see App. \ref{sec:theory_general}). The transmission/reflection is then extracted following Appendix \ref{app:3modes_trans} by solving the following coupled-modes equations
\begin{equation}\label{eq:coupled-modes-eqs}
\begin{split}
    \left( \partial_t^2 + \omega_{\lambda}^2 \right)\tilde{\mathcal{E}}_{\lambda} & = -\partial_t^2\sum_{k}\left(\tilde{w}_{\lambda, k}\tilde{P}_k^{\rm mp}+\tilde{u}_{\lambda, k}\tilde{P}_k^{\rm isb}\right),
    \\
     \left(\partial_t^2 +\bar{\omega}_B^2(k) \right)\tilde{P}_k^{\rm mp} & = N_{\rm qw}\omega_P^2 \sum_{\lambda}\tilde{w}_{\lambda, k}\tilde{\mathcal{E}}_{\lambda},
     \\
     \left(\partial_t^2 + \omega_{\rm isb}^2 \right)\tilde{P}_k^{\rm isb} & = N_{\rm qw}\omega_P^2 \sum_{\lambda}\tilde{u}_{\lambda, k}\tilde{\mathcal{E}}_{\lambda}.
\end{split}
\end{equation}
Here $\tilde{\mathcal{E}}_{\lambda} $ is the electric field amplitude of the $\lambda$-th cavity mode, with frequency $\omega_{\lambda}$; $\tilde{P}_k^{\rm mp}$ is the dielectric polarization amplitude of the MP $k$-th mode, with frequency $\bar{\omega}_B(k)$ which includes the $k$-dependent non-local Coulomb contribution; $\tilde{P}_k^{\rm isb}$ is the dielectric polarization amplitude of the ISB transition with frequency $\omega_{\rm isb}$.
As detailed in the Appendix \ref{app:pol_T_nonlocal} $\tilde{w}_{\lambda, k}, \tilde{u}_{\lambda, k}$ are coefficients that depend on the overlap between the $\lambda$-th spatial cavity mode and the $k$-th MP/ISB mode. These coefficients carry the fundamental information about the cavity-induced spatial inhomogeneity which is eventually able to trigger the observed non-locality.

It is worth stressing that, while the cavity modes are \emph{transverse} modes \cite{cohen1989photons} (whose dynamics are in the first line of Eq. \eqref{eq:coupled-modes-eqs}), we do not discard the longitudinal component of the electric field, which gives rise to what we call the \emph{Coulomb contribution} mentioned in Eq. \eqref{eq:E-field_inside}.
As indeed described in the previous section \ref{sec:coulomb}, and detailed in Appendix \ref{sec:theory_general}-\ref{app:3modes_trans}, it is this longitudinal contribution which provides a $k$-dependent depolarization shift affecting  the MP frequency, $\bar{\omega}_B(k)$, and emerges during the interaction with the TM mode.

We see a good agreement of the measured spectra in Fig. \ref{fig:figspectra}a with the calculated spectra in Fig. \ref{fig:figspectra}b. Comparing more closely in Fig. \ref{fig:figspectra}d, the calculated spectra accurately demonstrate interaction of the ISB polariton with the Coulomb-shifted and broadened MP. In addition, the calculation also models the blue-shift of the ISB polariton at low magnetic field, when the MP frequency is much smaller than the ISB one and $\omega_B\rightarrow 0$. 
This shift is compatible with the standard polariton gap due to ultrastrong coupling, and is present even though we don't observe clear polariton splitting due to the broadened cyclotron linewidth. However, since this shift is much smaller than the linewidth of the ISB polariton itself and on the order of our experimental system's spectral resolution, we cannot experimentally determine its quantitative value with accuracy. 

The TE mode is also well reproduced, showing a clear anti-crossing at the MP frequency, in contrast to the TM mode. We note for the calculation in \ref{fig:figspectra}b we use a value of 3.508 THz for the uncoupled TE cavity mode frequency, which is extracted from the experiment in Fig. \ref{fig:figspectra}a. This is slightly shifted from the value 3.8 THz of ab-initio simulations in Fig. \ref{fig:fig1}b, likely due to slightly mismatched cavity dimensions from their nominal values. This and other simulation parameters used can be found in Table \ref{tab:numerical_params} in Appendix \ref{sec:params}.

For comparison, in Appendix \ref{sec:withoutnonlocality} we also show the calculated spectra without including the Coulomb interaction, which reproduces the TE spectra, but does not accurately reproduce the experimental spectra for the TM mode. This demonstrates that to accurately model light-matter coupled systems with electronic transitions and inhomogeneous electric fields, it is crucial to include the influence of the field inhomogeneity on the Coulomb interactions in the 2DEG.  

\subsection{Hamiltonian Description and Ultra-Strong Coupling}

From the classical dielectric theory in Eq. \eqref{eq:coupled-modes-eqs} (see in Appendix \ref{sec:theory_general} for more details), we can also derive the respective Hamiltonian for the quantized system, which allows us to understand the dominant energies in the system.
Introducing the annihilation operators $a_{\lambda},b_k,c_k$, respectively for the $\lambda$-th cavity mode, MP and ISB $k$-th mode, the total system's Hamiltonian is given as the sum of four contributions $H = H_{c} + H_{\rm mp} + H_{\rm isb} + H_{\rm I}$, where (see Appendix \ref{app:Ham_form} for the derivation)
\begin{equation}
    \begin{split}
        H_{c} & = \sum_{\lambda}\hbar \omega_{\lambda}a_{\lambda}^{\dag} a_{\lambda} 
        \\
        H_{\rm mp} & = \sum_k \hbar\omega_B b^{\dag}_k b_k 
        \\
        H_{\rm isb} & = \sum_k \hbar\omega_{\rm isb} c^{\dag}_k c_k,
    \end{split}
\end{equation}
are, respectively, the cavity, MP, and ISB individual Hamiltonians.
The light-matter interaction Hamiltonian, describing the coupling between the electrodynamical cavity field and the MP, ISB excitations, is given by
\begin{equation}\label{eq:LM-int_HAM}
    \begin{split}
        &H_{\rm I} =
        \\
        &-i\frac{\hbar \tilde{\omega}_P}{2}\sum_{\lambda, k} \left(a_{\lambda} - a_{\lambda}^{\dag}\right)\left[ \tilde{w}_{\lambda, k}\left(b_{k} + b_{k}^{\dag}\right) + \tilde{u}_{\lambda, k}\left(c_{k} + c_{k}^{\dag}\right) \right]
        \\
        & + \sum_{\lambda}\frac{\hbar \tilde{\omega}_P^2}{4\omega_{\lambda}}\left[ \sum_k \left(\tilde{w}_{\lambda, k}\left(b_{k} + b_{k}^{\dag}\right) + \tilde{u}_{\lambda, k}\left(c_{k} + c_{k}^{\dag}\right)\right) \right]^2 
        \\
        & + \frac{\hbar \tilde{\omega}_P^2}{4\omega_B}\sum_k\zeta_k \left(b_{k} + b_{k}^{\dag}\right)^2.
    \end{split}
\end{equation}
Here we introduced the multiple-QW collective plasma frequency as $\tilde{\omega}_P = \sqrt{N_{\rm qw}}\omega_P$.

From Eq. \eqref{eq:LM-int_HAM} we see that the light-matter interaction is not given by a single term, but it is rather structured.
In the first line, we have the direct linear coupling between the cavity electric field and the total matter's polarization, obtained as a weighted sum of the MP and ISB polarizations: 
\begin{equation}\label{eq:Hlin}
\begin{split}
    H_{\rm lin} = & - i \frac{\hbar \tilde{\omega}_P}{2}\sum_{\lambda, k} \left(a_{\lambda} - a_{\lambda}^{\dag}\right) 
    \\
    & \times \left[ \tilde{w}_{\lambda, k}\left(b_{k} + b_{k}^{\dag}\right) + \tilde{u}_{\lambda, k}\left(c_{k} + c_{k}^{\dag}\right) \right].   
\end{split}
\end{equation}
This term scales linearly with the plasma frequency (the light-matter coupling constant), $\omega_P$, and, under the \emph{rotating-wave approximation} (RWA) gives the photon-matter flip-flop interaction that defines the polariton excitation \cite{ciuti_RevModPhys.85.299}.

In the second line of Eq. \eqref{eq:LM-int_HAM}, we have a matter-only term, scaling quadratically with the collective plasma frequency, $\tilde{\omega}_P^2$
\begin{equation}\label{eq:HP2-term}
    H_{P^2} = \sum_{\lambda}\frac{\hbar \tilde{\omega}_P^2}{4\omega_{\lambda}}\left[ \sum_k \left(\tilde{w}_{\lambda, k}\left(b_{k} + b_{k}^{\dag}\right) + \tilde{u}_{\lambda, k}\left(c_{k} + c_{k}^{\dag}\right)\right) \right]^2.
\end{equation}
This term is the famous $P^2$-term (polarization square), ensuring the stability of the system, and that also rules out any possible superradiant phase transition in the ground-state \cite{Todorov2010,todorovPhysRevB.85.045304,de_bernardis_cavity_2018, Rokaj_2018}. 

The last term in  Eq. \eqref{eq:LM-int_HAM} describes the Coulomb non-locality emerging from Eqs. \eqref{eq:E-field_inside}-\eqref{eq:omB_dispersion}-\eqref{eq:longwave_exp_zetak}
\begin{equation}\label{eq:Hnonloc}
H_{\rm nloc.} = \frac{\hbar\tilde{\omega}_P^2}{4\omega_B}\sum_k\zeta_k \left(b_{k} + b_{k}^{\dag}\right)^2,    
\end{equation}
The non-locality term is quadratic in the collective plasma frequency, $\tilde{\omega}_P^2$, and it can also be labelled as a $P^2$-term. 
However, contrary to the $P^2$-term in Eq. \eqref{eq:HP2-term}, which emerges from the transverse part of the electromagnetic interactions \cite{cohen1989photons,de_bernardis_cavity_2018, deBernardis2023}, the non-locality emerges from the longitudinal part of the electric field \cite{deBernardis2023}, and thus, while similar, these two terms are very different in nature and must not be confused.

The $\tilde{\omega}_P^2$, quadratic scaling of the non-locality term in Eq. \eqref{eq:Hnonloc}, clearly links its phenomenology with the so-called \emph{ultra-strong coupling} (USC) regime \cite{solano_RevModPhys.91.025005, FriskKockum2019, DeBernardis:24}. 
In this regime, the light-matter coupling strength becomes as large as the intrinsic cavity or QW energy scales (in our case, see Table \ref{tab:numerical_params}, $\hbar \tilde{\omega}_P \sim 1\,{\rm THz}\sim \hbar \omega_{\lambda}, \hbar \omega_{B}, \hbar\omega_{\rm isb}$).
In this way, the linear cavity-QW coupling given by Eq. \eqref{eq:Hlin} becomes sub-leading, indicating that, whenever the inhomogeneity of the system allows the cavity to excite $k>0$ modes (i.e. the $\tilde{w}_{\lambda, k}, \tilde{u}_{\lambda, k}$ overlap coefficients are non-zero at finite $k$), the USC regime is dominated by non-locality effects, affecting non-perturbatively the phenomenology of cavity QED light-matter interactions \cite{de_bernardis_cavity_2018}.

\subsection{Generality of Non-Locality}

For sufficiently doped quantum wells, the non-locality behavior only depends on the inhomogeneity of the cavity mode, and can thus be recovered in other types of heterostructure, and for different cavity geometries. 
To demonstrate this, in Figure \ref{fig:parabolic} we show the case of ISB polaritons in a cavity containing parabolic quantum wells \cite{Goulain2023}. 
Here, the ISBT is still coupled to the TM mode of the cavity, but the continuously-graded material composition of the quantum well allows for a lower frequency ISB transition, and therefore larger ISB polariton coupling strength. 
To match the TM mode frequency to the ISB frequency, the cavity dimensions are slightly different from the rectangular well case, with no GaAs buffer region (Fig. \ref{fig:parabolic}a). 
Despite the altered cavity dimensions, we still observe significant inhomogeneity in the fields, particularly at the interface between the metallic cavity plate and the gap. In addition, the lack of a buffer region means that the electrons in the quantum wells also experience the fields at the grating edges, which are highly inhomogeneous. 

As for the previous samples, Figure \ref{fig:parabolic}b shows the measured spectra when a magnetic field is applied, tuning the MP resonance across the ISB polaritons. Here, interaction of the MP with both polariton branches is visible thanks to the closer frequency alignment between the TM cavity mode and ISB, although the spectral features are slightly less clear due to the broadened polariton branches. For both polariton branches, these features include a cavity-induced broadening around the crossing point, a blue-shift of the MP from its bare frequency $\omega_B$ indicated as a black line as seen in the rectangular well case in Figure \ref{fig:figspectra}.

By demonstrating the effect of Coulomb interaction in different cavities with two different heterostructures, we highlight that the level of non-locality can be tailored through cavity design, independently from the quantum well system. To further support this, in Appendix \ref{sec:withoutnonlocality} we show the calculated spectra for both the cavities in Fig. \ref{fig:figspectra} and Fig. \ref{fig:parabolic} without including Coulomb interaction, demonstrating how the experimental spectra is clearly modified in the experiment by non-locality. Moreover, by further increasing the coupling strength either through different cavity design or increasing the electron density, the Coulomb interaction and thereby non-locality can be further enhanced.

\section{Conclusion}
\label{Conclusion}
In conclusion, we have experimentally demonstrated how the spatial inhomogeneity of a cavity mode can fundamentally alter the response of a two-dimensional electron gas (2DEG) in the strong light-matter coupling regime in the presence of a strong magnetic field. 
Using a multi-mode metal-insulator-metal (MIM) cavity containing quantum wells, we access different coupling regimes by tuning the the magnetoplasmon resonance as a function of magnetic field and selectively coupling it to distinct cavity or polaritonic modes, the latter resulting from the hybridization of cavity modes with the intersubband transition in the QW. 

We find that the MP–cavity coupling in a spatially inhomogeneous field results in significant new features, including spectral shifts attributable to Coulomb interaction. Furthermore, the breakdown of translational invariance - induced purely by the spatial profile of the cavity field - demonstrates a method for altering materials through cavity coupling that arises not from vacuum fluctuations but simply from the cavity field spatial profile. The cavity-induced Coulomb interactions can be activated on demand via magnetic field tuning and are controlled through the cavity design.

These findings demonstrate that, in order to accurately model light-matter coupled systems with electronic transitions and inhomogeneous electric fields, it is crucial to include the influence of the field inhomogeneity on the Coulomb interactions in the 2DEG.
Beyond this, the ability to manipulate polariton properties dynamically through cavity mode selection and magnetic field tuning opens new possibilities for tunable polaritonic devices, where the tripartite interplay can be also exploited to study the cavity protection phenomena or the quantum vacuum of ultrastrongly coupled systems \cite{deBernardis2022,deBernardis2023}.

\begin{acknowledgments}
We thank Gian Marcello Andolina, Bianca Turini, and Alberto Nardin for interesting and insightful discussions.
D.D.B. acknowledges funding from the European Union - NextGeneration EU, "Integrated infrastructure initiative in Photonic and Quantum Sciences" - I-PHOQS [IR0000016, ID D2B8D520, CUP B53C22001750006]. IC acknowledges support from the Provincia Autonoma di Trento; from the Q@TN Initiative; from the National Quantum Science and Technology Institute through the PNRR MUR Project PE0000023-NQSTI, co-funded by the European Union - NextGeneration EU.
J.F., G.S. and L.H. acknowledge funding through the SNF grant 200021-227521. This work was  also supported by the European Union Future and Emerging Technologies (FET) Grant No. 737017 (MIR-BOSE) and partially by the French RENATECH network.
\end{acknowledgments}

\section*{Data Availability}
The data supporting the findings of this article are openly available at \cite{ourdata}. 

\appendix

\section{General theoretical description}
\label{sec:theory_general}
Here we develop a fully classical description of the transmittivity/reflectivity of the THz cavity coupled to two-dimensional quantum well (QW).
The quantum well is characterized by a polarization density $\textbf{P}(\textbf{r})$, having the dimensions of dipole moment per volume.
The polarization density can be then linearly decomposed in all the contributions due to the independent transitions in the material.
While these transitions have fully quantum origin, it will be clear that a classical description fully matches the pure quantum one \cite{Ciuti2005,ciuti_PhysRevB.81.235303}.
Here for simplicity we focus on the case where the polarization is fully given by a MP transition \cite{Scalari2012}, but the generalization to other type of transitions (i.e. intersubband \cite{Ciuti2005}) is straightforward.

While the theory is developed in a fully general way, we will focus here on the simplest case of a planar, translationally invariant cavity filled with a medium having $\epsilon_r = 1$ with no grating.

\subsection{Maxwell equation with polarizable matter}
\label{sec:maxwell_eq_general}

We first start by considering Maxwell equations
\begin{equation}\label{eq:Maxwell_inhomogeneous_div}
		\vec{\nabla}\cdot \textbf{E} = \frac{\rho}{\epsilon_0}
\end{equation}
\begin{equation}\label{eq:Maxwell_inhomogeneous_rot}
		\vec{\nabla} \times \textbf{B} = \frac{1}{c^2} \left( \frac{\textbf{J}}{\epsilon_0} + \frac{\partial}{\partial t} \textbf{E} \right)
\end{equation}

\begin{equation}\label{eq:Maxwell_structural_div}
		\vec{\nabla}\cdot \textbf{B} = 0
\end{equation}
\begin{equation}\label{eq:Maxwell_structural_rot}
		\vec{\nabla} \times \textbf{E} = - \frac{\partial}{\partial t} \textbf{B}.
\end{equation}
As a first task in developing our theory we rewrite these equation in a minimal form to treat problems with polarizable matter.

Here, the charge density is fully given by the polarization density of the QW
\begin{equation}
    \rho = - \vec \nabla \cdot \textbf{P}_{3D},
\end{equation}
which is related to the current by the dipole (long wavelength) approximation
\begin{equation}
    \textbf{J} = \partial_t \textbf{P}_{3D}.
\end{equation}
Here we use the label $3D$ in $\textbf{P}_{3D}$ to distinguish from the two-dimensional QW polarization $\textbf{P}$ which will be introduced later on.

Combining Eq. \eqref{eq:Maxwell_inhomogeneous_rot} with the rotor of Eq. \eqref{eq:Maxwell_structural_rot} we obtain
\begin{equation}
    - c^2\nabla^2 \textbf{E} + c^2\vec \nabla (\vec \nabla\cdot \textbf{E}) = - \partial_t^2 \textbf{E} - \frac{1}{\epsilon_0} \partial_t^2 \textbf{P}_{3D}.
\end{equation}

To further reduce the system of equations and make it solvable, we need to take the Gauss law into account
\begin{equation}
    \vec{\nabla}\cdot \textbf{E} = -\frac{\vec \nabla \cdot \textbf{P}_{3D}}{\epsilon_0},
\end{equation}
so to isolate the so called transverse and longitudinal parts of the electric field
\begin{equation}
    \textbf{E} = \textbf{E}^{\parallel} + \textbf{E}^{\perp}.
\end{equation}
here
\begin{equation}\label{eq:E_parallel}
    \textbf{E}^{\parallel} = \frac{1}{\epsilon_0}\vec \nabla( G\star \vec \nabla \cdot \textbf{P}_{3D} )
\end{equation}
and
\begin{equation}
    \textbf{E}^{\perp} = \partial_t \textbf{A}.
\end{equation}
The vector $\textbf{A}$ is the vector potential, having the property $\vec{\nabla} \times \textbf{A} = \textbf{B}$ and $\vec{\nabla}\cdot \textbf{A} = 0$. As a consequence, $\vec \nabla \cdot \textbf{E}^{\perp} = 0$ by construction, coinciding with the standard definition of transverse vector \cite{cohen1989photons}.
We also introduced the Green's function of the Poisson equation, $G$, which solves $-\nabla^2G(\textbf{r}, \textbf{r}') = \delta(\textbf{r} - \textbf{r}')$ with metallic boundary conditions on the plates and zero electric potential difference between them. The $\star$ denotes the convolution operator in real space.
As a consequence also $\vec{\nabla}\times \textbf{E}^{\parallel}  = 0$ by construction.
These definitions are general and true in a cavity setup or any other confined geometry.

We finally arrive at the main equation describing the electric field dynamics in the grating cavity
\begin{equation}\label{eq:electric_field_equation_main}
    - c^2\nabla^2 \textbf{E}^{\perp} + \partial_t^2 \textbf{E}^{\perp} =  - \frac{1}{\epsilon_0} \partial_t^2 \left[ \textbf{P}_{3D} + \vec \nabla( G\star \vec \nabla \cdot \textbf{P}_{3D} )  \right].
\end{equation}
The term inside the square brackets on the left-hand side is the transverse projected polarization density
\begin{equation}
    \textbf{P}^{\perp}_{3D} = \textbf{P}_{3D} + \vec \nabla( G\star \vec \nabla \cdot \textbf{P}_{3D} ), 
\end{equation}
having the property
\begin{equation}
    \vec \nabla \cdot \textbf{P}^{\perp}_{3D} = 0.
\end{equation}
Its longitudinal projection is instead
\begin{equation}\label{eq:pol_parallel}
    \textbf{P}^{\parallel}_{3D} = - \vec \nabla( G\star \vec \nabla \cdot \textbf{P}_{3D} ).
\end{equation}
Evidently $\textbf{P}^{\perp}_{3D} + \textbf{P}^{\parallel}_{3D} = \textbf{P}_{3D}$.

We consider the transverse electric field decomposed on the cavity eigenmodes
\begin{equation}\label{eq:E-field_lambda_decompos}
    \textbf{E}^{\perp}(\textbf{r}, z) = \sum_{\lambda} \textbf{w}_{\lambda}(\textbf{r}, z)\mathcal{E}_{\lambda}(t).
\end{equation}
Here we use $\textbf{r}=(x,y)$ as the in-plane position, keeping explicit $z$, and $V$ is the system's volume. This convention will be especially convenient when we introduce the two-dimensional quantum well.
In particular, $\textbf{w}_{\lambda}(\textbf{r},z)$ is an adimensional function that fully characterises the $\lambda$ electromagnetic eigenmode.
The index $\lambda$ is intended as an generalized index, comprising the polarization index and eventual wavevector.
It has the properties that $\vec{\nabla}\cdot \textbf{w}_{\lambda}(\textbf{r},z) = 0$ and it satisfy the Helmholtz equation 
\begin{equation}
    -c^2\nabla^2\textbf{w}_{\lambda}(\textbf{r},z) = \omega_{\lambda}^2 \textbf{w}_{\lambda}(\textbf{r},z).
\end{equation}
These functions are orthonormal and normalized such that
\begin{equation}\label{eq:w_normalization}
    \int d^3 r \, \textbf{w}_{\lambda}(\textbf{r},z)\cdot \textbf{w}_{\lambda'}(\textbf{r},z) = \delta_{\lambda \lambda'},
\end{equation}
where $V$ is the system's volume.
Projecting both sides of Eq. \eqref{eq:electric_field_equation_main} on these eigenmodes, we finally obtain the main equation for the amplitude of the cavity modes
\begin{equation}\label{eq:E_field_amplitude}
    \partial_t^2 \tilde{\mathcal{E}}_{\lambda} + \omega_{\lambda}^2\tilde{\mathcal{E}}_{\lambda}  = \partial_t^2 \left[ \int d^2r\,dz \,\textbf{w}_{\lambda}(\textbf{r},z)\cdot \mathbf{P}_{3D}(\textbf{r},z) \right].
\end{equation}
Here $\tilde{\mathcal{E}}_{\lambda} = \epsilon \mathcal{E}_{\lambda}$ and has the same dimensional units of $\mathbf{P}_{3D}(\textbf{r},z) \sqrt{V}$.

Importantly in Eq. \eqref{eq:E_field_amplitude} the longitudinal polarization has disappeared because
\begin{equation}\label{eq:P_parallel_projection}
\begin{split}
    &\int d^2 rdz\,\textbf{w}_{\lambda}(\textbf{r},z)\cdot  \vec \nabla [ G\star \vec \nabla \cdot \textbf{P}_{3D} ](\textbf{r}, z) =  
    \\
    &\int d^2 r \, dz\,\textbf{w}_{\lambda}(\textbf{r},z)  \cdot \vec{\nabla}\left[ \varphi_{\textbf{P}}(\textbf{r},z) \right]  
    \\
    & = \int d^2 r \, dz\,\vec{\nabla}\cdot \left[ \textbf{w}_{\lambda}(\textbf{r},z)  \varphi_{\textbf{P}}(\textbf{r},z) \right] 
    \\
    &-   \int d^2 r \, dz\,\left[\vec{\nabla}\cdot \textbf{w}_{\lambda}(\textbf{r},z)\right] \varphi_{\textbf{P}}(\textbf{r},z)  
    \\
    &= 0.
\end{split}
\end{equation}
where 
\begin{equation}
    \varphi_{\textbf{P}}(\textbf{r},z) = \int d^2r' dz' G(\textbf{r},z, \textbf{r}',z')\vec{\nabla}'\cdot\textbf{P}_{3D}(\textbf{r}',z').
\end{equation}
The term in the third line of Eq. \eqref{eq:P_parallel_projection} is a boundary term, vanishing in a infinitely large system.
The term in the fourth line is zero as well by definition of the transverse mode functions $\vec{\nabla}\cdot\textbf{w}_{\lambda}(\textbf{r},z) = 0$.

\subsection{Long wavelength Green's function}
\label{sec:Green}

The most important object to compute the longitudinal contribution to the dielectric response is the Green's function of the Poisson equation
\begin{equation}
    -\nabla^2G(\textbf{r},z,\textbf{r}', z') = \delta^{(2)}(\textbf{r} - \textbf{r}')\delta(z-z'),
\end{equation}
this gives rise to the well-known Coulomb interaction.

In free space it gives the standard Coulomb potential $G(\textbf{r},z,\textbf{r}', z') = 1/(4\pi|\textbf{r}_{3D} - \textbf{r}_{3D}'|)$, where $\textbf{r}_{3D} = (\textbf{r}, z)$, while in a infinite parallel mirrors cavity one must includes all the image charges extra terms \cite{vladimirov1971equations}.

For our purposes, we don't actually need the real space representation, but rather its Fourier k-space one.
To have the maximally accurate description, we would need to consider the solution relative to the grating patch cavity, with appropriate boundary conditions.
In practice, this is quite challenging, so we restrict our discussion to the simplified case where two parallel, perfectly metallic mirrors form the cavity.
The distance between the two mirrors is the cavity length, and is called $L_c$.
Introducing $k=\sqrt{k_x^2+k_y^2}$, the Poisson equation is rewritten as
\begin{equation}
    \left( -\partial_z + k^2 \right) G_k (z, z') = \delta(z-z'),
\end{equation}
imposing the metallic boundary conditions $G_k (z=0, z')=G_k (z=L_c, z')=0$.
The solution is given by
\begin{equation}
     G_k (z, z') = g_k(z,z')\Theta(z-z') + g_k(z',z)\Theta(z'-z),
\end{equation}
where 
\begin{equation}
\label{eq:gkzz}
\begin{split}
    &g_k(z,z') = \frac{e^{-k|z-z'|}}{2k} 
    \\
    &- \frac{1}{2k\sinh(kL_c)}\left[ \sinh (kz)e^{k(z'-L_c)} + \sinh (k(L_c-z)) e^{-kz'} \right],    
\end{split}
\end{equation}
and $\Theta(z)$ is the Heaviside step function.

The long-wavelength expansion for $L_ck\to 0$ gives
\begin{equation}
    \lim_{k\rightarrow 0} g_k(z,z') =  \frac{1}{2k} - \frac{z z'}{L_c}.
\end{equation}
It is worth stressing that this expansion holds only if $k\ll 1/L_c$. Having a box of size $L_x$, for which $k=2\pi/L_x$, this immediately implies $L_x \gg L_c$, suggesting that the limit $L_c\to\infty$ does not commute with the longwavelength limit.

Taking $z=z'=L_c/2$, we have
\begin{equation}
    k^2G_k(L_c/2, L_c/2) \approx \frac{k}{2}-\frac{L_ck^2}{4}
\end{equation}
Instead, to calculate the Green's function in the free space limit (no cavity), we have to take the limit $L_c\to\infty$ already in \eqref{eq:gkzz}, which recovers the standard Coulomb dispersion,
\begin{equation}
    k^2G_k(L_c/2, L_c/2) \approx \frac{k}{2}.
\end{equation}

\subsection{Cyclotron polarization and non-locality}
\label{app:cycl}
While the cyclotron (or MP) transition in a 2DEG (two-dimensional electron gas) is a fully quantum phenomenon, due to harmonicity, it can be equivalently described by a classical system.
In particular, here we develop a coarse-grained description of the MP polarization, which is also often dubbed in the literature as macroscopic dielectric theory.

Let's consider a single electron as a classical charged particle, confined in a two dimensional plane at $z=0$, in a strong homogeneous perpendicular magnetic field $\textbf{B}_{\rm ext} = (0,0,B_{\rm ext})$ and in-plane electric field $\textbf{E}(\textbf{r},z)=(E_x(\textbf{r},z), E_y(\textbf{r},z), 0)$. As before we take as a convention $\textbf{r}=(x,y)$ as the in-plane position.
Calling $\delta \textbf{r}$ the in-plane coordinate of the electron, the equations of motion are
\begin{equation}
m\,\delta \ddot{\textbf{r}} = e \delta \dot{\textbf{r}}\times \textbf{B}_{\rm ext} + e\textbf{E}(\delta \textbf{r}, z=0).
\end{equation}
By integrating out the equation for $\delta y$, we obtain the equation for $\delta x$ (the same applies to $\delta y$)
\begin{equation}\label{eq:single_e_cycl_x}
    \delta\ddot{x} + \omega_B^2\delta x = \frac{e}{m}E_x(\delta \textbf{r}, 0),
\end{equation}
which describes an harmonic motion coupled to the electric field.

We then extend the description to many electrons $n=1,2\ldots \delta N_e$, all localized around the position $\textbf{r}$. 
Here we completely neglect the interaction between these electron. While this assumption is surely wrong in a real physical system, here is motivated by the fact that we can interpret $\delta \textbf{r}_n$ as the coordinate of effective degrees of freedom, resulting from a proper quantum treatment where the Fermionic nature of electrons allows to remap the interacting problem in a independent particle model \cite{lipparini_book_2008}.

We take a long-wavelength approximation approximating the electric field as homogeneous around that position $\textbf{E}(\delta \textbf{r}_1, \delta \textbf{r}_2, \ldots \delta \textbf{r}_N) \approx \textbf{E}(\textbf{r})$.
We identify the material total dipole moment at this fixed position as $e\sum_n \delta \textbf{r}_n$.
From Eq. \eqref{eq:single_e_cycl_x} we obtain the equation of the oscillating total dipole 
\begin{equation}
    \partial_t^2 \sum_n \delta x_n + \omega_B^2 \sum_n \delta x_n = \frac{e\delta N_e}{m}E_x(\textbf{r}).
\end{equation}
Here, the displacements $\delta x_n$ are relative to the fixed position $\textbf{r}$, which can be interpreted as the center of mass position of this electronic ensemble.
In general, we now understand that the macroscopic two-dimensional polarization can be defined as 
\begin{equation}
\textbf{P}(\textbf{r}) = e\sum_n \frac{\delta \textbf{r}_n}{\delta S }
\end{equation}
where $\delta S$ is a small surface element containing the $\delta N_e$ electrons \cite{jackson_classical_2013}.
A similar treatment applies to the more general three-dimensional polarization $\textbf{P}_{3D}$ defined in the previous subsection, but here, for simplicity, we focus directly on the specific case of our interest.
Since the general electrodynamics equations developed in Sec. \ref{sec:maxwell_eq_general} require a three-dimensional polarization, it is worth noticing that the two-dimensional one can be casted to a three-dimensional one via delta function
\begin{equation}\label{eq:pol_3d_magnetoplasm}
    \textbf{P}_{3D}(\textbf{r},z) = \textbf{P}(\textbf{r})\delta(z).
\end{equation}

We then arrive at the main equation describing the harmonic motion of the in-plane polarization
\begin{equation}\label{eq:pol_dyn_general}
    \partial_t^2 \textbf{P}(\textbf{r}) + \omega_B^2 \textbf{P}(\textbf{r}) = \omega_P^2 L_c \epsilon \textbf{E}(\textbf{r}, 0).
\end{equation}
Here we have introduced the plasma frequency as
\begin{equation} 
    \omega_P = \sqrt{\frac{e^2 n_{2D}}{\epsilon m L_c}},
\end{equation}
where we have used the three-dimensional electron density as $n_{3D} = \delta N_e/\delta V$, leading to the two-dimensional density $n_{2D} = n_{3D}L_c$.
$\epsilon = \epsilon_r \epsilon_0$ is the dielectric permittivity of the considered material while $L_c$ is the cavity height.
From here on the system's volume is given by $V=L_cS$. A full quantum treatment leads to a different coupling constant, which depends explicitly on the filling factor $\nu$ \cite{ciuti_PhysRevB.81.235303}. However, our classical description gives very similar quantitative results without the complexity of a full quantum description and with the advantage of having a more transparent interpretation.

In light of the development in the previous subsection, we further manipulate Eq. \eqref{eq:pol_dyn_general} to account for the split between transverse and longitudinal electric field components.
We use $\textbf{E}(\textbf{r}, 0) = \textbf{E}^{\perp}(\textbf{r}, 0) + \vec \nabla [ G\star \vec \nabla \cdot \textbf{P}_{3D} ](\textbf{r}, 0)/\epsilon = \textbf{E}^{\perp}(\textbf{r}, 0) - \textbf{P}^{\parallel}(\textbf{r})/\epsilon$, which is a consequence of Eqs. \eqref{eq:E_parallel}-\eqref{eq:pol_parallel}.
The main equation of motion for the cyclotron polarizability is then
\begin{equation}\label{eq:pol_2d_withCoulomb}
    \partial_t^2 \textbf{P}(\textbf{r}) + \omega_B^2\textbf{P}(\textbf{r}) = \omega_P^2\left(\epsilon L_c\textbf{E}^{\perp}(\textbf{r}, 0) - \textbf{P}^{\parallel}(\textbf{r})\right).
\end{equation}
The two-dimensional version of the longitudinal polarization appearing in this equation is rewritten explicitly as
\begin{equation}
    \textbf{P}^{\parallel}(\textbf{r}) = L_c\int d^2 r' \vec{\nabla} \vec{\nabla}' G(\textbf{r}, z=0, \textbf{r}', z'=0)  \cdot \textbf{P}(\textbf{r}').
\end{equation}
Here we have integrated by parts and discarded the boundary terms.

Eq. \eqref{eq:pol_2d_withCoulomb} is particularly important because it shows how the Coulomb force enters the polarization dynamics through the longitudinal part of the field. This term is what is often referred to as \emph{non-locality} \cite{hopfield_nonlocal_PhysRev.132.563, Rajabali2021, Monticone_2025}, and we will see that it is the major factor responsible for depolarization shift phenomena in non-homogeneous MP excitations. To this aim it is worth to express Eq. \eqref{eq:pol_2d_withCoulomb} in Fourier space, by considering $\tilde{\textbf{P}}_{\textbf{k}} = \int d^2r/S\, e^{-i\textbf{k}\cdot \textbf{r}} \textbf{P}(\textbf{r})$ (same for $\textbf{E}^{\perp}(\textbf{r}, 0)$).
We have that
\begin{equation}\label{eq:pol_2d_withCoulomb}
    \partial_t^2 \tilde{\textbf{P}}_{\textbf{k}} + \omega_B^2\tilde{\textbf{P}}_{\textbf{k}} = \omega_P^2\left(\epsilon L_c \tilde{\textbf{E}}_{\textbf{k}} - L_c\,\textbf{k}\,\tilde{G}_{\textbf{k}}\textbf{k}\,\cdot \tilde{\textbf{P}}_{\textbf{k}} \right).
\end{equation}
Here we have assumed that the system is translation invariant in the plane, so that $G(\textbf{r}, z=0, \textbf{r}', z'=0) = G(\textbf{r}-\textbf{r}', z=0, z'=0)$.
Restricting for simplicity the system only the $x$-dimension and considering that $\lim_{k\rightarrow 0}\tilde{G}_{\textbf{k}} = 1/(2k)$ \cite{Takae_polar_green_2013,de_bernardis_cavity_2018} we have
\begin{equation}
    \partial_t^2 \tilde{P}_k + \bar{\omega}_B^2(k) \tilde{P}_k = \omega_P^2\epsilon L_c \tilde{E}_k,
\end{equation}
where the cyclotron frequency is now shifted by the $k$-dependent depolarization shift \cite{Rajabali2021}
\begin{equation}
    \bar{\omega}_B^2(k) = \omega_B^2 + \omega_P^2\frac{L_ck}{2}.
\end{equation}

More generally, we can write
\begin{equation}\label{eq:non-locality}
    \bar{\omega}_B^2(k, z) = \omega_B^2 + \omega_P^2\,\zeta_k(z),
\end{equation}
where 
\begin{equation}
    \zeta_k = \lim_{k\rightarrow 0} k^2\tilde{G}_k(z=z').
\end{equation}

\subsection{Polaritonic transmission in the presence of non-locality}
\label{app:pol_T_nonlocal}
Combining Eq. \eqref{eq:E_field_amplitude} with Eq. \eqref{eq:pol_2d_withCoulomb}, we can finally address the coupled light-matter system, where the cyclotron MP excitation is hybridized to the polaritonic/cavity modes to form new hybridized modes.

In order to account for the polaritonic origin of the TM mode we must also include the ISBT polarization, splitting the total polarization in Eq. \eqref{eq:E_field_amplitude} into $\textbf{P}_{3D} = \textbf{P}_{3D}^{\rm mp} +  \textbf{P}_{3D}^{\rm isb}$ \cite{deBernardis2023}.

Again, restricting to only the $x$-direction, and taking a Fourier transform with respect to time, we obtain the coupled modes equations
\begin{equation}\label{eq:coupled_modes1}
    \left(\omega_{\lambda}^2 - \omega^2\right)\tilde{\mathcal{E}}_{\lambda} = \omega^2\sum_{k}\left(\tilde{w}_{\lambda, k}\tilde{P}_k^{\rm mp}+\tilde{u}_{\lambda, k}\tilde{P}_k^{\rm isb}\right),
\end{equation}
\begin{equation}\label{eq:coupled_modes2}
    \left(\bar{\omega}_B^2(k) - \omega^2 \right)\tilde{P}_k^{\rm mp} = \omega_P^2 \sum_{\lambda}\tilde{w}_{\lambda, k}\tilde{\mathcal{E}}_{\lambda},
\end{equation}
\begin{equation}\label{eq:coupled_modes3}
    \left(\omega_{\rm isb}^2 - \omega^2 \right)\tilde{P}_k^{\rm isb} = \omega_P^2 \sum_{\lambda}\tilde{u}_{\lambda, k}\tilde{\mathcal{E}}_{\lambda},
\end{equation}
Here we have shifted the electric field in Eq. \eqref{eq:E-field_lambda_decompos} to
\begin{equation}
    \tilde{\mathcal{E}}_{\lambda} \mapsto \tilde{\mathcal{E}}_{\lambda}\sqrt{V}/L_c.
\end{equation}
In this way, the coupling matrices are given by the dimensionless quantities (by restoring the full two-dimensionality dependence)
\begin{equation}
    \tilde{w}_{\lambda, \textbf{k}} = \sqrt{V}\int \frac{d^2 r}{S} e^{-i\textbf{k}\cdot \textbf{r}} \textbf{w}_{\lambda}(\textbf{r},0)\cdot \textbf{u}_x,
\end{equation}
\begin{equation}
    \tilde{u}_{\lambda, \textbf{k}} = \sqrt{V}\int \frac{d^2 r}{S} e^{-i\textbf{k}\cdot \textbf{r}} \textbf{w}_{\lambda}(\textbf{r},0)\cdot \textbf{u}_z,
\end{equation}
where $\textbf{u}_x = (1,0,0), \textbf{u}_z = (0,0,1) $ are the $x$ or $z$ unit pointers.
Moreover, we have introduced the ISBT frequency, which is assumed to be $k$-independent and not affected by non-locality.

The transmission/reflection can then be obtained through the formula \cite{deBernardis2022,deBernardis2023}
\begin{equation}
    \mathcal{T} = \sum_{\lambda}\gamma_{\lambda}\omega_{\lambda} \sum_{\lambda '}\mathcal{M}^{-1}_{\lambda \lambda '} A_{\lambda '}.
\end{equation}
Here we introduced the photon losses of each mode $\gamma_{\lambda}$, which is incorporated in Eqs. \eqref{eq:coupled_modes1}-\eqref{eq:coupled_modes2} by shifting the poles $\omega_{\lambda}^2 - \omega^2\mapsto\omega_{\lambda}^2 - \omega^2 - i\gamma_{\lambda}\omega$ for each mode.
The dynamical matrix is defined as
\begin{equation}\label{eq:dyn_matrix_general}
\begin{split}
     \mathcal{M}_{\lambda \lambda '} = & \left(\omega_{\lambda}^2 - \omega^2\right)\delta_{\lambda \lambda '} - \omega^2 \omega_P^2\sum_k \frac{\tilde{u}_{\lambda, k} \tilde{u}_{\lambda ', k}}{\omega_{\rm isb}^2 - \omega^2}
     \\
     &- \omega^2 \omega_P^2\sum_k \frac{\tilde{w}_{\lambda, k} \tilde{w}_{\lambda ', k}}{\bar{\omega}_B^2(k) - \omega^2}.
\end{split}
\end{equation}
Here we also introduce losses: for the cyclotron excitation, $\kappa$, by substituting $\bar{\omega}_B^2(k) - \omega^2\mapsto\bar{\omega}_B^2(k) - \omega^2 - i\kappa\omega$, for the cavity, $\gamma$, by substituting $\omega_{\lambda}^2 - \omega^2\mapsto\omega_{\lambda}^2 - \omega^2 - i\gamma\omega$, for the ISBT $\kappa_{\rm isb}$, by substituting $\omega_{\rm isb}^2 - \omega^2\mapsto\omega_{\rm isb}^2 - \omega^2 - i\kappa_{\rm isb}\omega$.
The adimensional amplitudes $A_{\lambda}$ are instead given by the projections of the in-plane external drive amplitude $\textbf{I}(\mathbf{r})$ on the cavity eigenmodes 
\begin{equation}
    A_{\lambda} \sim  \frac{\int d^2r\, \textbf{w}_{\lambda}(\textbf{r}, z=L_c)\cdot \textbf{I}(\textbf{r})}{\sqrt{\int d^2r\, |\textbf{I}(\textbf{r})|^2 }}
\end{equation}

\section{Three-modes cavity theory}
\label{app:3modes_trans}

Here we specialize the transmissivity/reflectivity theory developed in Sec. \ref{sec:theory_general} to describe the experimental setup described in the main text, with the parameters derived in Sec. \ref{sec:params}.

We restrict the description to only the main two TM, TE cavity modes described before, and we can neglect the coupling between different cavity modes and essentially only keep the diagonal terms of $\mathcal{M}$ in Eq. \eqref{eq:dyn_matrix_general}.
Notice that the TM mode is a polaritonic mode, split into lower-polariton (LP) and upper-polariton (UP), so in the end we will see three modes in the transmission, TM$_{\rm LP}$, TM$_{\rm UP}$, TE, all hybridized with the MP.
The transmissivity can then be written as
\begin{equation}
    \mathcal{T}\approx \sum_{\lambda}\frac{ \gamma_{\lambda}\omega_{\lambda}A_{\lambda}}{\omega_{\lambda}^2 - \omega^2\epsilon_{\lambda}(\omega ) -i\gamma_{\lambda}\omega }
\end{equation}
where we introduced the ISBT and MP relative permittivities per mode $\epsilon_{\lambda}(\omega ) = 1+\chi^{\rm isb}_{\lambda}(\omega)+\chi^{\rm mp}_{\lambda}(\omega)$, determined by the ISBT and MP susceptibility
\begin{equation}
    \chi_{\lambda}^{\rm isb}(\omega) = \omega_P^2\sum_k \frac{|\tilde{u}_{\lambda, k} |^2 }{\omega_{\rm isb}^2 - \omega^2 - i\kappa_{\rm isb} \omega},
\end{equation}
\begin{equation}
    \chi_{\lambda}^{\rm mp}(\omega) = \omega_P^2\sum_k \frac{|\tilde{w}_{\lambda, k} |^2 }{\bar{\omega}_B^2(k) - \omega^2 - i\kappa \omega}.
\end{equation}

\begin{figure}
    \centering
    \includegraphics[width=\columnwidth]{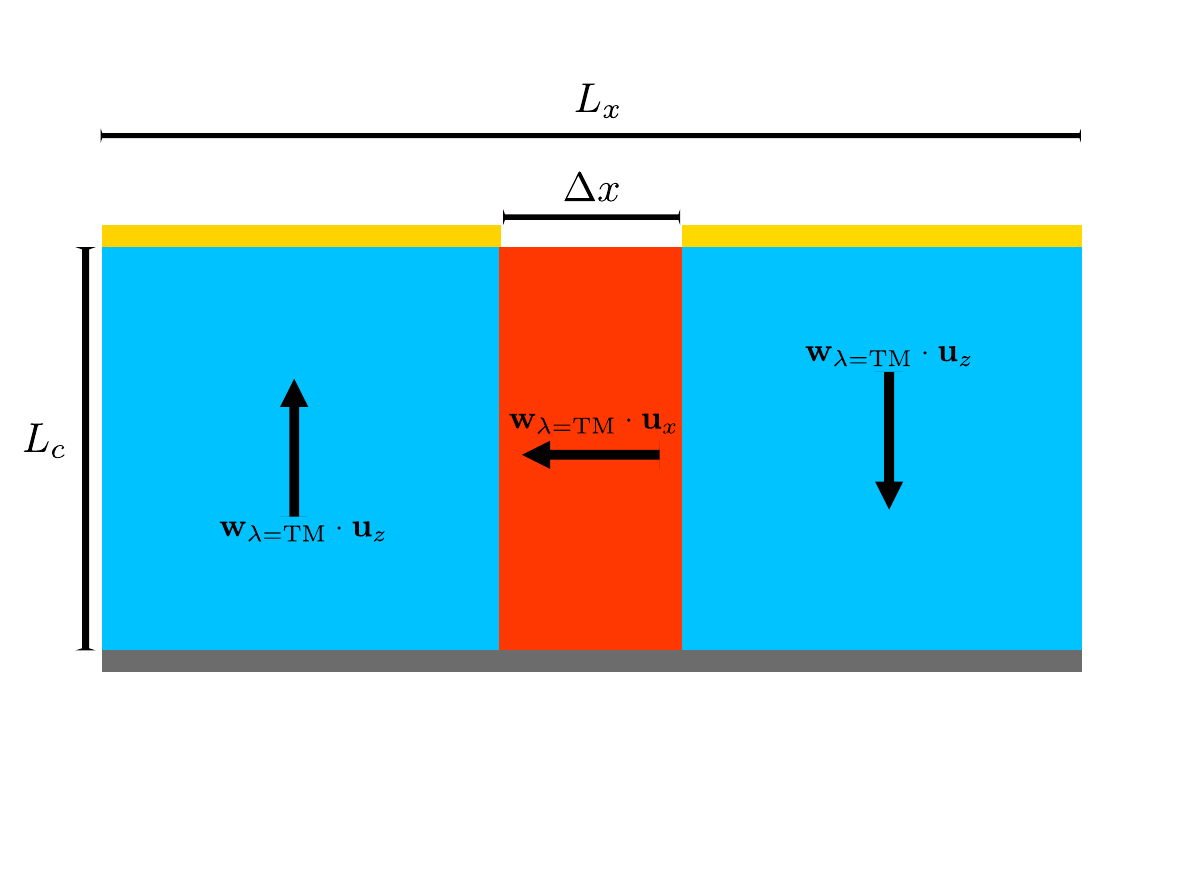}
    \caption{Schematic of the TM-modes spatial field polarization.}
    \label{fig:approx_modes}
\end{figure}

For the TM mode, we assume that the eigenmode $x$-projection is a box function, which is constant corresponding to the grating gap between the metallic patches, and zero in correspondence to the patch
\begin{equation}
    \textbf{w}_{\lambda={\rm TM}}(x,y,0) \cdot \textbf{u}_x \approx \sqrt{f\frac{1}{2\Delta x}} \Pi(x) \cos(k_{1}x).
\end{equation}
Here $\Pi(x)= [1-\Theta(x-\Delta x/2)]\Theta(x+\Delta x/2)$ is the box function and $\Theta(x)$ is the Heaviside step function.
The adimensional factor $f<1$ accounts for the non-optimal overlap between the QW and the cavity modes.
Moreover, the wavevectors $k_{1}$ are given by the periodicity of the structure, and we have that $k_{1}\approx 2\pi/L_x$.
This approximation is schematically represented in Fig. \ref{fig:approx_modes}.
Here, $L_x$ is the $x$-size of the quantum well.
Notice that the normalization factor $\sqrt{1/\Delta x}$ follows from the normalization condition of the eigenmodes in Eq. \eqref{eq:w_normalization}, taking the $y$-projection vanishing and the $z$-projection complementary to the $x$-one.
Their Fourier transform gives
\begin{equation}\label{eq:w}
\begin{split}
\tilde{w}_{\lambda={\rm TM}, k}
&= \sqrt{f\frac{\Delta x}{8L_x}}
\left[ {\rm sinc}\!\left((k-k_{1})\frac{\Delta x}{2}\right) \right.\\
&\quad \left. +\, {\rm sinc}\!\left((k+k_{1})\frac{\Delta x}{2}\right) \right]
\end{split}
\end{equation}

Here ${\rm sinc}(kx) = \sin (kx)/(kx)$.
For the $z$-projection instead, we only keep the $k=0$ term, since the ISBT is not substantially affected by non-locality.
We then have $\tilde{u}_{\lambda={\rm TM}, k} \approx \sqrt{f_{\rm isb} }\delta_{0k}$, where $f_{\rm isb}  < 1$.

For the TE mode, instead, we assume that it is completely constant across the plane so to have
\begin{equation}
    \textbf{w}_{\lambda={\rm TE}}(x,y,0) \cdot \textbf{u}_x \approx \sqrt{\frac{f}{L_x}},
\end{equation}
and $\textbf{w}_{\lambda={\rm TE}}(x,y,0) \cdot \textbf{u}_z \approx 0$.
For the MP, its normalized Fourier transform gives
\begin{equation}
    \tilde{w}_{\lambda={\rm TE}, k} = \delta_{k\,0}\sqrt{f},
\end{equation}
and $\tilde{u}_{\lambda={\rm TE}, k} = 0$ for the ISB.


We also assume that all the $N_{\rm qw}=57$ QWs couple in the same way to the cavity, and contribute equally to the Coulomb non-locality. We can thus include the factor $\sqrt{N_{\rm qw}}$ as a collective enhancement of the plasma frequency.
We have that the two modes MP susceptibility is now simplified to
\begin{equation}\label{eq:chiTM_approx}
\begin{split}
    \chi_{\lambda={\rm TM}}^{\rm mp}(\omega) &\approx f\frac{N_{\rm qw}\omega_P^2}{2}\frac{\Delta x}{L_x}\sum_{k>0} \frac{{\rm sinc}^2((k-k_1)\Delta x/2) }{\bar{\omega}_B^2(k) - \omega^2 - i\kappa \omega}
    \\
    \chi_{\lambda={\rm TE}}^{\rm mp}(\omega) &\approx f\frac{N_{\rm qw}\omega_P^2}{\omega_B^2 - \omega^2 - i\kappa \omega},
\end{split}
\end{equation}
while the ISBT gets 
\begin{equation}
\begin{split}
    \chi_{\lambda={\rm TM}}^{\rm isb}(\omega) &\approx f_{\rm isb}\frac{N_{\rm qw}\omega_P^2}{2} \frac{1 }{\omega_{\rm isb}^2 - \omega^2 - i\kappa_{\rm isb} \omega}
    \\
    \chi_{\lambda={\rm TE}}^{\rm isb}(\omega) &\approx 0.
\end{split}
\end{equation}
It is worth noticing that in Eq. \eqref{eq:chiTM_approx} we neglected the cross contribution proportional to ${\rm sinc}((k-k_1)\Delta x/2){\rm sinc}((k+k_1)\Delta x/2)$, which is a valid approximation if $\Delta x$ is sufficiently large, so that the overlap between these terms is negligible.
Then, we can resum the $\pm k_1$ contributions, rewriting the formula limited to only the $k>0$ modes.

In Eq. \eqref{eq:non-locality} of $\bar{\omega}_B^2(k)$, we must also use $N_{\rm qw}\omega_P^2$ instead of $\omega_P^2$.
This accounts for dipole-dipole near-field coupling between the slabs \cite{deBernardis2023}, for which the MP mode is actually a  superradiant collective mode emerging from coupled motion of all the quantum wells together.
This effective description holds well under the long-wavelength condition, for which $k\Delta z \ll 1$. Here $k$ is the typical MP wavelength and $\Delta z$ is the separation between the quantum wells along the $z$ axis. Since $k\sim 0-1/\Delta x$ we have that the inter-well separation must be $\Delta z \ll 10 \mu$m, which is surely satisfied.

To better visualize the effect of the non-locality (as in Fig. \ref{fig:figspectra}b), we can define an MP trajectory in the $B$-field, that interpolates between the bare cyclotron resonance $\omega_B$ and the brightest k-mode of the MP interacting with the TM cavity mode. 
Indeed, looking at Eq. \eqref{eq:w}, we realize that $k_1$ is the brightest mode when the MP is resonant with the TM cavity mode, while out of resonance the dominant MP mode is at $k=0$.
We then introduce a $B$-dependent $k$ wavevector
\begin{equation}
    k(B) = k_1 \frac{\gamma_{\rm TM}^2}{\left(\omega_{\rm TM} - \bar{\omega}_B(k_1)\right)^2+\gamma_{\rm TM}^2}.
\end{equation}
The red line MP trajectory in Fig. \ref{fig:figspectra}b is then simply given by $\omega_{\rm MP}(B) = \bar{\omega}_B(k(B))$, calculated on the parameters in the next Section \ref{sec:params}.
Naturally, this $\omega_{\rm MP}(B)$ trajectory is not rigorous, and the interpolation can be done with any other function rather than the Lorentzian used here. However, we find that it correctly visualizes the MP behavior.

\section{Parameter estimation}\label{sec:params}
Here we focus on the estimation for the parameters needed in the theory.
Since the QW consists in a slab of GaAs, we take $\epsilon \approx 13 \epsilon_0$ and $m\approx 0.067m_e$, where $\epsilon_0\approx 8.85\,{\rm pF/m}\approx 53.1\,e/(\mu{\rm V\, m})$ is the vacuum permittivity, $e$ and $m_e$ are the electron charge and mass.
The three most important quantities in this work are
\begin{equation}
\begin{split}
    \frac{e^2}{\epsilon_0}& \approx 1.8\times 10^4\,{\rm meV}\,{\rm nm},
    \\
    \frac{\hbar^2}{m} & \approx 1136\,{\rm meV}\,{\rm nm^2},
    \\
    \mu_B &= \frac{e\hbar^2}{2m} \approx 0.86 \frac{{\rm meV}}{{\rm T}}.
\end{split}
\end{equation}

Assuming a 2D electron density $n_{2D}\approx 10^{11}{\rm cm}^{-2} = 10^{-3}\,{\rm nm}^2$ we have that $\hbar \omega_P \approx 0.4\,{\rm meV}$, corresponding to 
\begin{equation}
    \frac{\omega_P}{2\pi}\approx 100 \, {\rm GHz.}    
\end{equation}
(we remind the energy-to-frequency conversion is given by $2\pi\hbar \approx 4.13\,{\rm meV/THz}$).

Considering the collective enhancement due to the presence of many quantum wells, for instance $N_{\rm qw}= 57$ and $n_{2D}\approx 3\times 10^{-11}$, as reported in the main text, we have 
\begin{equation}
    \frac{\sqrt{N_{\rm qw}}\omega_P}{2\pi} \approx 1.3 \, {\rm THz}.
\end{equation}

This value is quite large and it would give a much bigger Rabi splitting than what is observed on the TE mode in the main text $\Omega_R/(2\pi) \approx 600\,$GHz.
However, accounting for the smaller overlap between the QW and the cavity field, $f<1$ returns the correct value $\Omega_R/(2\pi) \approx 2f\sqrt{N_{\rm qw}}\omega_P/(2\pi)$.

From the experimental data we observe the MP to cross the TE mode at around $B\approx 8.5\,$T. 
Considering that $\omega_B = 2\mu_B B$, from this value we obtain
\begin{equation}
    \frac{\omega_B}{2\pi} \approx 3.5\,{\rm THz},
\end{equation}
indicating that the electron's mass is correctly $m* = 0.067 m_e$.

From the independent observation of the MP resonance linewidth, we obtain
\begin{equation}
    \frac{\kappa}{2\pi}\approx 300\,{\rm GHz}.
\end{equation}

The two main cavity modes are given by $\lambda=\lbrace{{\rm TM, TE}\rbrace}$, called transverse magnetic, and transverse electric, based on their polarization properties \cite{jackson_classical_2013}.
Both the square-well and parabolic well devices illustrated in the main text support these modes, due to their MIM, metallic-grating patch-cavity design.

\begin{table}[ht]
    \centering
    \begin{tabular}{ c | c | c }
         & Square Well & Parabolic Well\\
         $\omega_{\rm TM}/(2\pi)$ & 2.408 THz  & 1.5 THz\\
         $\omega_{\rm TE}/(2\pi)$ & 3.508 THz & 3.5 THz \\
         $\omega_{\rm isb}/(2\pi)$ & 2.73 THz & 1.4 THz \\
         $\sqrt{N_{\rm qw}} \omega_P/(2\pi)$ & 1 THz & 1.4 THz\\
         $f$ & 0.3 & 0.3 \\
         $f_{\rm isb}$ & 0.3 & 0.4 \\
         $\Delta x$ & 10 $\mu$m & 6 $\mu$m \\
         $L_c$ & 11.5 $\mu$m & 4 $\mu$m \\
         $L_x$ & 30$\mu$m & 30 $\mu$m\\
         $\gamma/(2\pi)$ & 50 GHz & 50 GHz\\
         $\kappa/(2\pi)$ & 300 GHz & 300 GHz \\
         $\kappa_{\rm isb}/(2\pi)$ & 200 GHz & 300 GHz\\ 
    \end{tabular}
    \caption{Parameters used for the numerical simulations in the main text}
    \label{tab:numerical_params}
\end{table}
All the other parameters, such as cavity linewidth, ISBT frequencies etc... are taken phenomelogically from the experimental data, and are consistently benchmarked with the standard literature.
The parameters used in the numerical simulations reported in the main text are summarized in Table \ref{tab:numerical_params}.
Notice that the cavity linewidth is take equal for all the modes $\gamma_{\lambda} = \gamma$.

\section{Comparison to simulation without the longitudinal Coulomb contribution}\label{sec:withoutnonlocality}

\begin{figure*}
    \centering
    \includegraphics[width= 16cm]{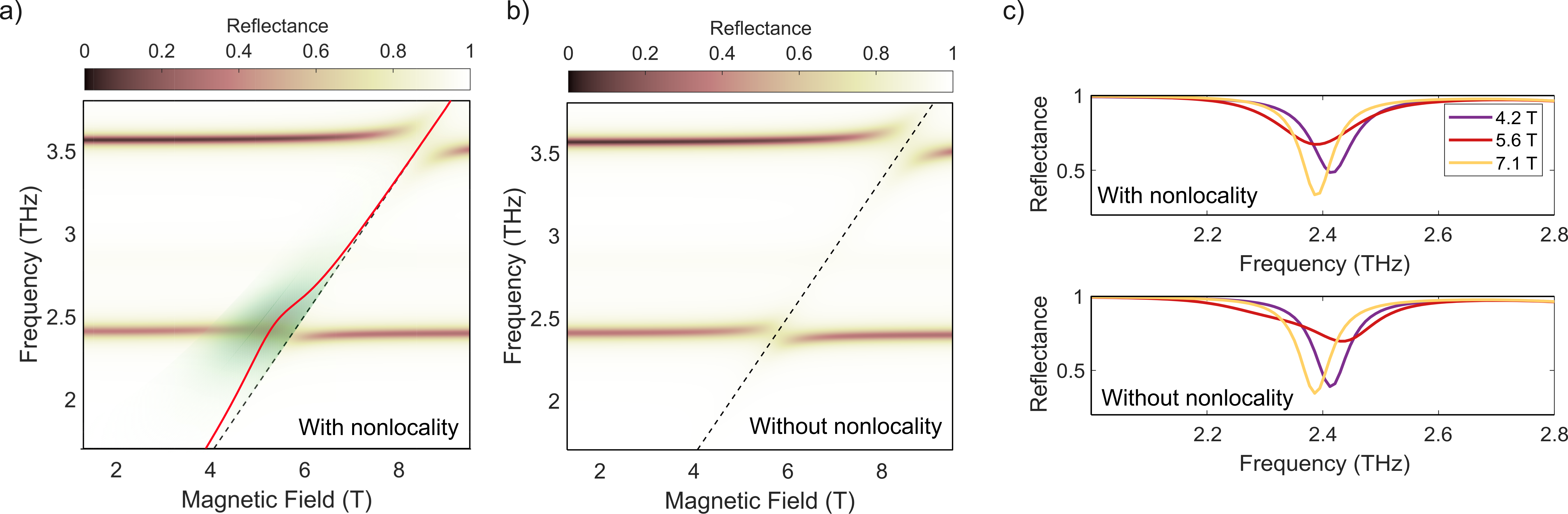}
    \caption{a) Same simulation as in Fig. \ref{fig:figspectra}. Red solid line is a guide for the eye obtained by interpolating between the bare cyclotron frequency and the brightest, blue-shifted, cyclotron k-mode (see Appendix \ref{app:3modes_trans} for further details). The green shaded region represents a collection of the various trajectories $\bar{\omega}_B(k)$ for increasing values of $k\in (0, \infty)$; the opacity of each line is given $|w_{\rm TM, k}|^2\gamma^2/( (\omega_{\rm TM}- \bar{\omega}_B(k))^2+\gamma^2 )$ accordingly to Eq. \eqref{eq:w}. b) Same simulation as in (a) but with $\bar{\omega}_B(k) = \omega_B$. c) Individual spectra of TM LP for three values of the magnetic field extracted from (a)-top and (b)-bottom, as in Fig. \ref{fig:figspectra}.}
    \label{fig:S1}
\end{figure*}

\begin{figure}
    \centering
    \includegraphics[width= 8cm]{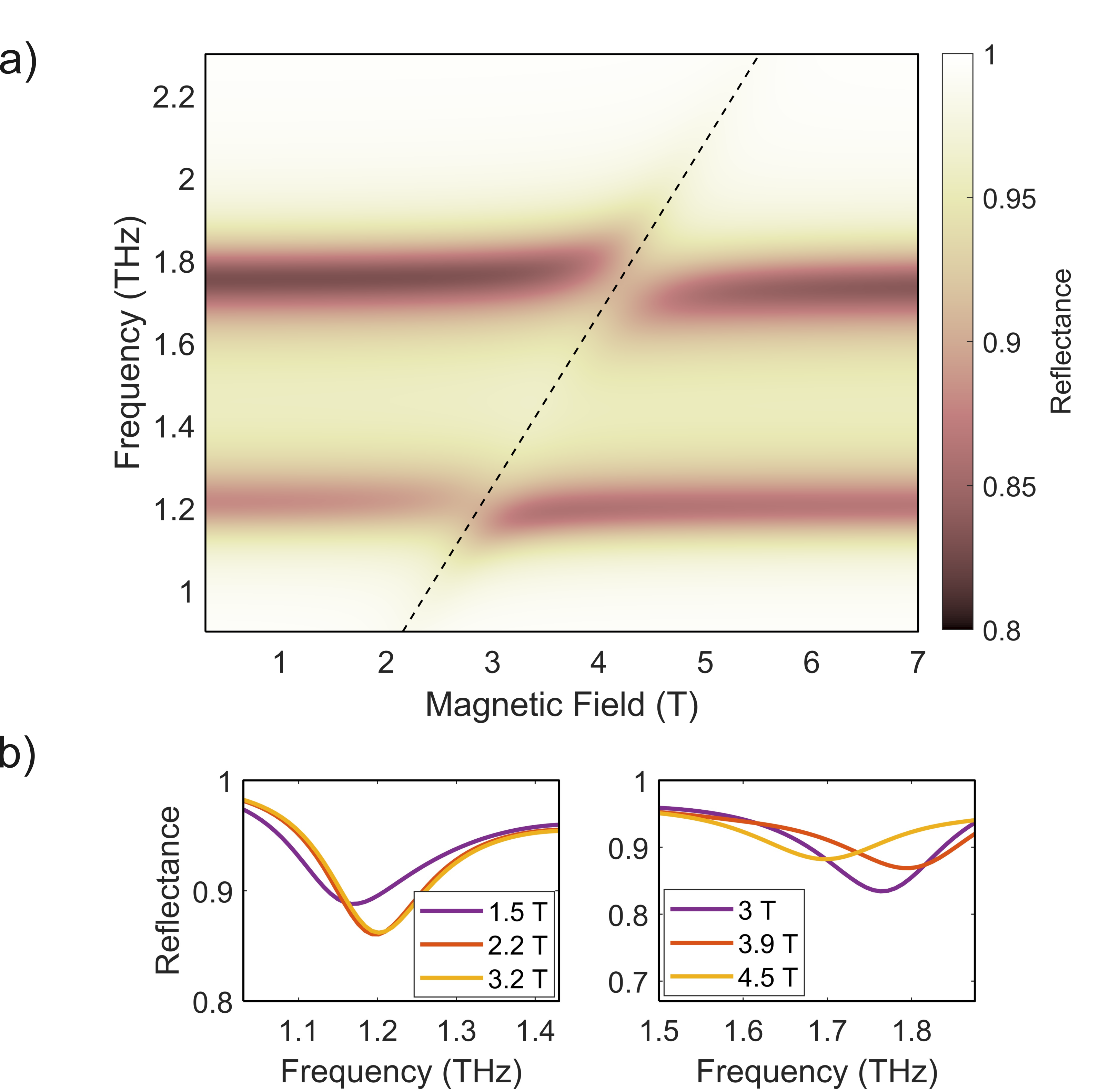}
    \caption{a) Same simulation as in Fig. \ref{fig:parabolic}, with $\bar{\omega}_B(k) = \omega_B$. b) Individual spectra of lower (left) and upper (right) polaritons for three values of the magnetic field extracted from (a), as in Fig. \ref{fig:parabolic}.}
    \label{fig:S2}
\end{figure}

In this section we further demonstrate the effect of non-locality by comparing our calculated spectra in Figures \ref{fig:figspectra} and \ref{fig:parabolic} to the case where we artificially remove the shift induced by Coulomb effects, setting $\bar{\omega}_B(k) = \omega_B$. In Fig. \ref{fig:S1} we show this comparison for the RQW case (Fig. \ref{fig:figspectra}). Without the inclusion of the k-dependent shift due to the longitudinal Coulomb interaction, Rabi splitting between the MP and the polariton lines is more evident and the interaction is perfectly symmetric about the cyclotron resonance frequency. For completeness, in Fig. \ref{fig:S2} we also show the calculated spectra without non-locality for the PQW case (for comparison to Fig. \ref{fig:parabolic}d.
\vspace{0.2 cm}

\vspace{0.2 cm}

\section{Hamiltonian formulation}
\label{app:Ham_form}

Here we consider the finalized coupled modes equations in Eqs. \eqref{eq:coupled_modes1}-\eqref{eq:coupled_modes2}-\eqref{eq:coupled_modes3} and we recast this system into the  Hamiltonian framework.
For simplicity here we omit the tilde $\tilde{\cdot}$ from all the variables.

First we introduce the two auxiliary parameters $\varepsilon_{\rm aux}, M_{\rm aux}$, defined such that
\begin{equation}
    N_{\rm qw}\omega_P^2 = \frac{1}{\varepsilon_{\rm aux}M_{\rm aux}}.
\end{equation}
Then we introduce a new variable for the electromagnetic field
\begin{equation}
    \mathcal{E}_{\lambda} = \varepsilon_{\rm aux}\partial_t A_{\lambda},
\end{equation}
and for the polarization
\begin{equation}
    P_{\lambda} = \sum_{k}\left(\tilde{w}_{\lambda, k}\tilde{P}_k^{\rm mp}+\tilde{u}_{\lambda, k}\tilde{P}_k^{\rm isb}\right).
\end{equation}

Then Eq. \eqref{eq:coupled_modes1} is rewritten as
\begin{equation}
    \partial_t^2A_{\lambda} = - \omega_{\lambda}^2A_{\lambda} - \partial_t \frac{P_{\lambda}}{\varepsilon_{\lambda}}.
\end{equation}
Introducing the canonical variable 
\begin{equation}\label{eq:Dcanonical}
    D_{\lambda} = \varepsilon_{\rm aux}\partial_t A_{\lambda} + P_{\lambda}
\end{equation}
so we rewrite the equation as the first Hamilton equation
\begin{equation}
    \partial_t D_{\lambda} = - \omega_{\lambda}^2A_{\lambda} = -\frac{\partial H_{\mathcal{E}}}{\partial A_{\lambda}},
\end{equation}
where the electric field Hamiltonian is introduced as
\begin{equation}
    H_{\mathcal{E}} = \sum_{\lambda}\frac{\left(D_{\lambda} - P_{\lambda}\right)^2}{2\varepsilon_{\rm aux}} + \frac{\epsilon_{\rm aux}\omega_{\lambda}^2}{2}A_{\lambda}^2.
\end{equation}
By construction also the second Hamilton equation is satisfied
\begin{equation}
    \partial_t A_{\lambda} = \frac{\partial H_{\mathcal{E}}}{\partial D_{\lambda}} = \frac{D_{\lambda}-P_{\lambda}}{\varepsilon_{\rm aux}},
\end{equation}
due to the definition given in Eq. \eqref{eq:Dcanonical}.

In a similar way we derive the Hamiltonian for the MP and ISB polarization equations.
Let's introduce the canonical moment of the MP/ISB degree of freedom
\begin{equation}
    \Pi_k^{\rm mp/isb} = M_{\rm aux} \partial_t P_k^{\rm mp/isb}.
\end{equation}
Eqs. \eqref{eq:coupled_modes2}-\eqref{eq:coupled_modes3} are then rewritten as a first Hamilton equation
\begin{equation}
\begin{split}
    \partial_t \Pi_k^{\rm mp/isb} &= -\frac{\partial H_{\rm mp/isb}}{\partial P_k^{\rm mp/isb}} -\frac{\partial H_{\mathcal{E}}}{\partial P_k^{\rm mp/isb}}    
    \\
    & = - M_{\rm aux}\omega_{\rm mp/isb}^2P_k^{\rm mp/isb} + \sum_{\lambda}w_{k\lambda}\frac{D_{\lambda}-P_{\lambda}}{\varepsilon_{\rm aux}},
\end{split}
\end{equation}
where
\begin{equation}
    H_{\rm mp} = \sum_k \frac{\left(\Pi_k^{\rm mp}\right)^2}{2M_{\rm aux}} + \frac{M_{\rm aux}\bar{\omega}_B^2(k)}{2}\left(P_k^{\rm mp}\right)^2
\end{equation}
\begin{equation}
    H_{\rm isb} = \sum_k \frac{\left(\Pi_k^{\rm isb}\right)^2}{2M_{\rm aux}} + \frac{M_{\rm aux}\omega_{\rm isb}^2}{2}\left(P_k^{\rm isb}\right)^2.
\end{equation}

The total Hamiltonian of the system is then
\begin{equation}
    H = H_{\mathcal{E}} +  H_{\rm isb} + H_{\rm mp}.
\end{equation}

Introducing the complex variables $a_{\lambda}$, $b_{k}$, $c_k$, respectively for the cavity, the MP and the ISB, we define
\begin{equation}
    \begin{split}
        D_{\lambda} & = -i \sqrt{\frac{\varepsilon_{\rm aux}\omega_{\lambda}}{2}}\left(a_{\lambda} - a_{\lambda}^{\dag}\right),
        \\
        A_{\lambda} & = \sqrt{\frac{1}{2\varepsilon_{\rm aux}\omega_{\lambda}}}\left(a_{\lambda} + a_{\lambda}^{\dag}\right),
    \end{split}
\end{equation}

\begin{equation}
    \begin{split}
        \Pi_{k}^{\rm mp} & = -i \sqrt{\frac{M_{\rm aux}\omega_{B}}{2}}\left(b_{k} - b_{k}^{\dag}\right),
        \\
        P_{k}^{\rm mp} & = \sqrt{\frac{1}{2M_{\rm aux}\omega_{B}}}\left(b_{k} + b_{k}^{\dag}\right),
    \end{split}
\end{equation}

\begin{equation}
    \begin{split}
        \Pi_{k}^{\rm isb} & = -i \sqrt{\frac{M_{\rm aux}\omega_{\rm isb}}{2}}\left(c_{k} - c_{k}^{\dag}\right),
        \\
        P_{k}^{\rm isb} & = \sqrt{\frac{1}{2M_{\rm aux}\omega_{\rm isb}}}\left(c_{k} + c_{k}^{\dag}\right).
    \end{split}
\end{equation}

The Hamiltonian $H = H_{\mathcal{E}} + H_{\rm mp} + H_{\rm isb}$ becomes
\begin{equation}
    \begin{split}
        H_{\mathcal{E}} & = \sum_{\lambda}\omega_{\lambda}a_{\lambda}^{\dag} a_{\lambda} 
        \\
        &- i \frac{\sqrt{N_{\rm qw}}\omega_P}{2}\sum_{\lambda, k} \left(a_{\lambda} - a_{\lambda}^{\dag}\right)\left[ \tilde{w}_{\lambda, k}\left(b_{k} + b_{k}^{\dag}\right) + \tilde{u}_{\lambda, k}\left(c_{k} + c_{k}^{\dag}\right) \right]
        \\
        & + \sum_{\lambda}\frac{N_{\rm qw}\omega_P^2}{4\omega_{\lambda}}\left[ \sum_k \tilde{w}_{\lambda, k}\left(b_{k} + b_{k}^{\dag}\right) + \tilde{u}_{\lambda, k}\left(c_{k} + c_{k}^{\dag}\right) \right]^2,
    \end{split}
\end{equation}
\begin{equation}
    \begin{split}
        H_{\rm mp} & = \sum_k\left[ \omega_B b^{\dag}_k b_k + \frac{N_{\rm qw}\omega_P^2}{4\omega_B}\zeta_k \left(b_{k} + b_{k}^{\dag}\right)^2 \right]
        \\
        H_{\rm isb} & = \sum_k \omega_{\rm isb} c^{\dag}_k c_k.
    \end{split}
\end{equation}
By reintroducing $\hbar$ and imposing the right commutation relation one has immediately also the quantized three-partite polaritonic system.

\bibliographystyle{apsrev4-2}  

\end{document}